\documentclass[fleqn,usenatbib]{mnras}

\usepackage{newtxtext,newtxmath}

\usepackage[T1]{fontenc}

\DeclareRobustCommand{\VAN}[3]{#2}
\let\VANthebibliography\thebibliography
\def\thebibliography{\DeclareRobustCommand{\VAN}[3]{##3}\VANthebibliography}

\usepackage{graphicx}	
\usepackage{amsmath}	

\title[Chrono-chemo-dynamics of unusual stars]{The odd bunch: chrono-chemo-dynamics of sixteen unusual stars from Kepler}

\author[Arthur Alencastro Puls et al.]{
Arthur Alencastro Puls,$^{1,2,7}$\thanks{E-mail: AlencastroPuls@iap.uni-frankfurt.de (AAP)}
Luca Casagrande,$^{2,7}$
Stephanie Monty,$^{3,2,7}$
David Yong,$^{2,7}$
Fan Liu,$^{4,7}$\newauthor
Dennis Stello$^{5,6,7}$
and Mikkel N. Lund$^{6}$
\\
$^{1}$Goethe University Frankfurt, Institute for Applied Physics, Max-von-Laue-Str. 12, 60438, Frankfurt am Main, Germany\\
$^{2}$Research School of Astronomy and Astrophysics, Australian National University, Canberra, ACT 2611, Australia\\
$^{3}$Institute of Astronomy, Madingley Road, Cambridge, CB3 0HA, UK\\
$^{4}$School of Physics \& Astronomy, Monash University, Clayton VIC 3800, Australia\\
$^{5}$School of Physics, University of New South Wales, Sydney, NSW 2052, Australia\\
$^{6}$Stellar Astrophysics Centre (SAC), Department of Physics and Astronomy, Aarhus University, Ny Munkegade 120, DK-8000 Aarhus C, Denmark\\
$^{7}$ARC Centre of Excellence for All Sky Astrophysics in 3 Dimensions (ASTRO3D), Australia
}

\date{Accepted 2023 May 02. Received 2023 April 27; in original form 2022 December 22}

\pubyear{2023}

\begin{document}
\label{firstpage}
\pagerange{\pageref{firstpage}--\pageref{lastpage}}
\maketitle

\begin{abstract}
In this study we combine asteroseismic, spectroscopic and kinematic information to perform a detailed analysis of a sample of 16 stars from the Kepler field. Our selection focuses on stars that appear to contradict Galactic chemical evolution models: young and $\alpha$-rich, old and metal-rich, as well as other targets with unclear classification in past surveys. Kinematics are derived from Gaia DR3 parallaxes and proper motions, and high-resolution spectra from HIRES/Keck are used to calculate chemical abundances for over 20 elements. This information is used to perform careful checks on asteroseismic masses and ages derived via grid-based modelling. Among the seven stars previously classified as young and $\alpha$-rich, only one seems to be an unambiguously older object masking its true age. We confirm the existence of two very old ($\geq$11~Gyr), super metal rich ($\geq$0.1~dex) giants. These two stars have regular thin disc chemistry and in-plane solar circle orbits which fit well in the picture of radial migration via the churning mechanism. The alternative explanation that these stars have younger ages would require mass-loss rates which strongly increases with increasing metallicity. Finally, we suggest further investigations to explore the suitability of Zn as a chemical clock in red giants.
\end{abstract}

\begin{keywords}
Galaxy: abundances -- Galaxy: kinematics and dynamics -- stars: fundamental parameters
\end{keywords}



\section{Introduction}

One of the well-established facts in nuclear astrophysics is that most chemical elements found in nature are synthesised inside stars \citep[][]{Burbidge:1957aa,Nomoto:2013aa,Tatischeff:2018,Kobayashi:2020}. That, along with the fact that stars largely conserve chemical information about their cosmic cradles in their atmospheres, means that we may use stellar abundances to probe the evolution of the Milky Way (hereafter, the Galaxy) and, in the bigger picture, of the Universe \citep[][]{Freeman:2002aa,Dotter:2017aa}.

In the past decade the advent of space based asteroseismology has allowed us to enrich the field of Galactic Archaeology by enabling measurements of masses and ages of a large number of red giants with unprecedented level of precision. Surveys such as those conducted by \citet[][]{Yu:2018} have provided asteroseismic information for more than 16,000 objects. As a consequence, now it is possible to examine Galactic evolution models in a more detailed manner \citep[e.g.,][]{Stello:2015,Zinn:2022}, in particular their information with respect to temporal evolution.

It has been observed that the Galactic disc has a spatial-temporal pattern regarding stellar abundances. Stars formed in the inner part of the disc tend to be more metal-rich than those formed in the outer disc at similar age \citep[e.g.,][]{Schonrich:2009aa,Bensby:2011aa,Casagrande:2011aa,Casagrande:2016aa}, and spatial dependence in the separation between high- and solar-$\alpha$ abundances is observed as well \citep[e.g.,][]{Queiroz:2020}. Also, as a natural consequence of stellar nucleosynthesis, younger stars are expected to be more metal-rich than their older counterparts in the same Galactic environment. This is expected assuming a simplified 'closed-box' evolution \citep[][]{Matteucci:2021}, but also in more refined scenarios \citep[e.g.,][]{Spitoni:2019}. However, there is no clear age-metallicity relation in the solar neighbourhood \citep[][]{Edvardsson:1993aa,Casagrande:2011aa}.

It has been proposed that the absence of an age-metallicity relation among stars in the solar neighbourhood is a consequence of stellar radial migration \citep[][]{Sellwood:2002,Haywood:2008,Roskar:2008}. That is also an explanation for the presence of super metal-rich stars in the solar circle \citep[e.g.,][and references therein]{Trevisan:2011}. On the other hand, the ages of some of these super metal-rich stars might be inaccurate \citep[as commented by][their ages for Red Clump stars must be taken with caution]{Pinsonneault:2018}, a refinement of stellar evolution models may be required, leading to an improvement in the determination of fundamental stellar parameters. For instance, a better understanding of the Red Giant Branch (RGB) mass loss rate might improve our understanding of stars in the subsequent evolutionary stage, the Red Clump (RC).

Since the advent of \emph{Kepler} \citep[][]{Koch:2010}  and \emph{CoRoT} \citep[][]{Baglin:2006aa}, red giants identified as young and $\alpha$-rich have been extensively studied \citep[][]{Martig:2015aa,Chiappini:2015aa,Yong:2016aa,Jofre:2016aa,Matsuno:2018aa,SilvaAguirre:2018,Hekker:2019,Zhang:2021,Jofre:2023}. These objects are interesting because the ratio between $\alpha$-elements and Fe is expected to decrease with the onset of Type-Ia Supernovae, which occurs relatively fast compared to the age of the Galactic disc \citep[][]{Matteucci:2001,Kilic:2017}. Many of the studies dissecting these anomalous $\alpha$-rich giants have suggested that their true ages have been masked by events of mass accretion. Hence, their current masses would be larger than their initial masses. Still, a detailed study may reveal more information about these interesting targets.

In that sense, our aim in this paper is to analyse a set of stars from the Galactic disc that have been previously identified as potential outliers with respect to chemical evolution models. We make use of the finest spectroscopic, asteroseismic, and astrometric information available to date for that set of targets. In Section~\ref{sec:targetsandobservations} we summarise the sample under study, and in Section~\ref{sec:analysis} we describe the methodology. Section~\ref{sec:results} contains the results and discussion. Our final remarks are presented in Section~\ref{sec:conclusions}.

\section{Targets and observations}
\label{sec:targetsandobservations}

\begin{table*}
	\centering
	\caption{Program stars, their adopted average seismic parameters and sources: (A) \citet{Yu:2018}, (B) \citet{Pinsonneault:2018}, (C) \citet{Casagrande:2014aa}, (D) This work. Parallaxes from Gaia DR2 and DR3 have zero-point calibrations from \citet{Zinn2019} and \citet{Lindegren:2021}, respectively.}
	\label{tab:tab1}
	\begin{tabular}{rrrrrrrr}
		\hline
		KIC & RA & Dec & $\Delta \nu$ & $\nu_{\mathrm{max}}$ & Src & $\varpi_{\mathrm{DR2}}$ & $\varpi_{\mathrm{DR3}}$ \\
		& (J2000) & (J2000) & $\mu$Hz & $\mu$Hz & & mas & mas \\
        \hline
               KIC2845610 & 19 19 27.4059 & +38 01 41.196 &  7.069 $\pm$ 0.052 &   91.65 $\pm$   4.66 & A & 1.5878 $\pm$ 0.0292 & 1.6516 $\pm$ 0.0110 \\
               KIC3455760 & 19 37 45.7037 & +38 35 35.637 &  4.860 $\pm$ 0.012 &   48.32 $\pm$   0.50 & A & 1.0835 $\pm$ 0.0275 & 1.0080 $\pm$ 0.0131 \\
               KIC3833399 & 19 02 43.0468 & +38 54 59.335 &  4.146 $\pm$ 0.024 &   37.12 $\pm$   0.72 & A & 1.8780 $\pm$ 0.0236 & 1.7855 $\pm$ 0.0109 \\
               KIC5512910 & 18 55 30.9211 & +40 42 44.672 &  4.337 $\pm$ 0.030 &   38.24 $\pm$   0.71 & A & 0.2931 $\pm$ 0.0207 & 0.2571 $\pm$ 0.0138 \\
               KIC5707338 & 19 29 34.5475 & +40 54 18.624 &  6.421 $\pm$ 0.037 &   81.89 $\pm$   1.28 & A & 1.0415 $\pm$ 0.0317 & 1.0614 $\pm$ 0.0220 \\
               KIC6605673 & 19 26 41.3873 & +42 03 41.578 & 68.010 $\pm$ 0.950 & 1263.00 $\pm$  49.00 & C & 3.0358 $\pm$ 0.0254 & 2.9965 $\pm$ 0.0130 \\
               KIC6634419 & 19 54 48.4403 & +42 04 29.868 &  7.238 $\pm$ 0.031 &   92.39 $\pm$   2.11 & A & $\cdots$            & 0.9982 $\pm$ 0.1695 \\
               KIC6936796 & 19 11 30.4158 & +42 28 39.546 &  3.196 $\pm$ 0.044 &   22.36 $\pm$   0.82 & A & 0.9443 $\pm$ 0.0206 & 0.8636 $\pm$ 0.0113 \\
               KIC6940126 & 19 16 27.6755 & +42 28 23.477 & 21.610 $\pm$ 2.050 &  256.30 $\pm$   1.30 & D & 0.9992 $\pm$ 0.0164 & 0.9475 $\pm$ 0.0132 \\
               KIC7595155 & 19 09 10.6703 & +43 16 18.998 &  3.879 $\pm$ 0.044 &   28.95 $\pm$   0.73 & A & 0.7125 $\pm$ 0.0225 & 0.6411 $\pm$ 0.0097 \\
               KIC8145677 & 18 50 43.8808 & +44 03 14.383 &  4.513 $\pm$ 0.128 &   30.54 $\pm$   0.34 & A & 0.7115 $\pm$ 0.0201 & 0.6105 $\pm$ 0.0103 \\
               KIC9002884 & 18 54 05.7764 & +45 20 47.467 &  0.885 $\pm$ 0.029 &    4.70 $\pm$   0.01 & B & 0.3859 $\pm$ 0.0203 & 0.2738 $\pm$ 0.0128 \\
               KIC9266192 & 18 57 05.5191 & +45 43 28.972 &  6.317 $\pm$ 0.034 &   78.30 $\pm$   5.94 & A & 1.6266 $\pm$ 0.0235 & 1.5940 $\pm$ 0.0119 \\
               KIC9761625 & 19 09 38.0164 & +46 35 25.277 &  1.429 $\pm$ 0.023 &    9.23 $\pm$   0.41 & A & 0.3996 $\pm$ 0.0228 & 0.3036 $\pm$ 0.0111 \\
              KIC10525475 & 19 10 21.3464 & +47 43 19.394 &  4.335 $\pm$ 0.023 &   39.12 $\pm$   0.91 & A & 0.9129 $\pm$ 0.0186 & 0.8508 $\pm$ 0.0100 \\
              KIC11823838 & 19 45 52.9235 & +50 02 30.463 &  4.508 $\pm$ 0.025 &   42.70 $\pm$   0.80 & A & 0.8634 $\pm$ 0.0269 & 0.7343 $\pm$ 0.0176 \\
              
	    \hline
	\end{tabular}
\end{table*}

The high-resolution spectral data analysed in this paper were observed with the High Resolution Echelle Spectrometer \citep[HIRES,][]{Vogt:1994} at the W. M. Keck Observatory, in the same campaign as the spectra studied by \citet[][hereafter Paper~I]{AlencastroPuls:2022}. Data reduction was performed in the same method as in Paper~I, using the Mauna Kea Echelle Extraction (\textsc{makee}) and \textsc{iraf}.

\begin{figure}
	\includegraphics[width=\columnwidth]{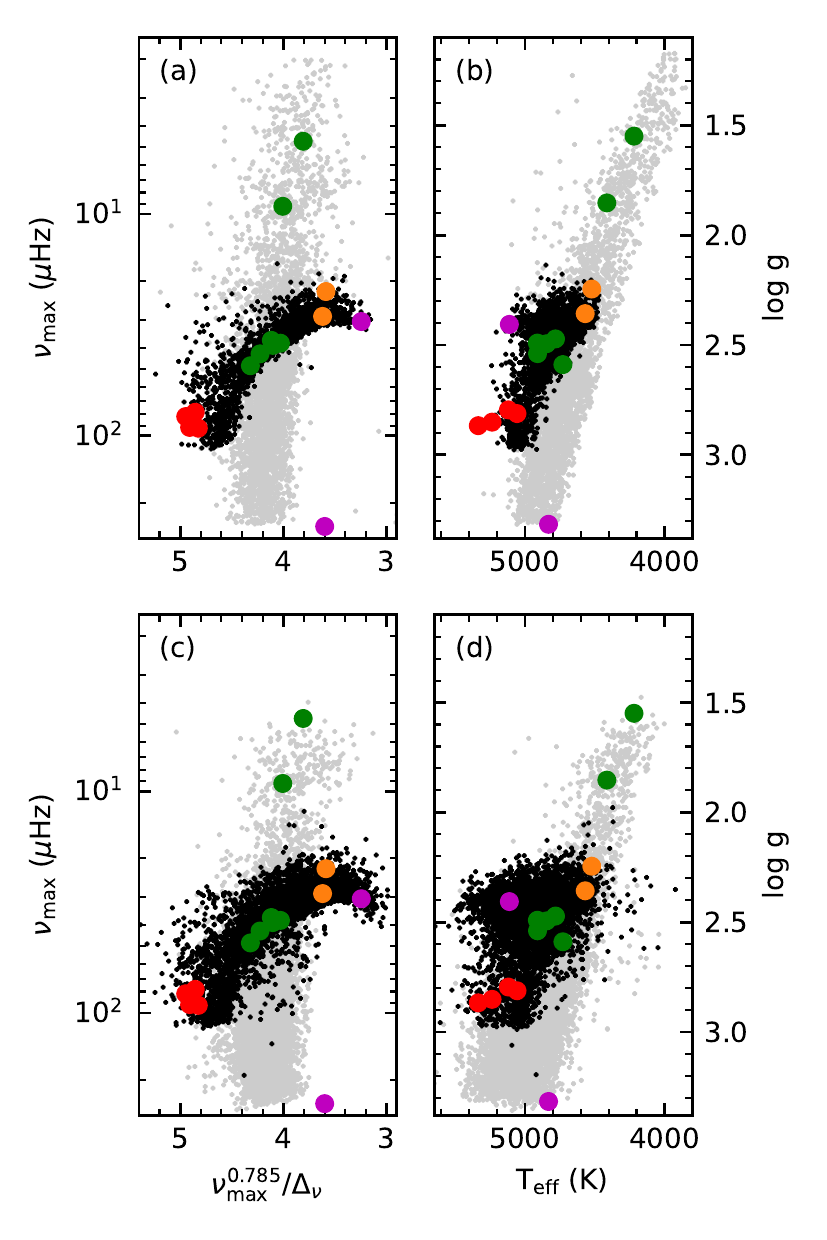}
    \caption{How our sample of red giants (circles) compares to the datasets from APOKASC-2 \citep[][ panels (a) and (b)]{Pinsonneault:2018} and \citet[][panels (c) and (d)]{Yu:2018}.
    Black points are stars classified as core He-Burning in their works, while the grey points represent stars ascending the RGB.
    Circles' colours represent the four groups enumerated in Section~\ref{sec:targetsandobservations}.
    \emph{Green}: APOKASC young $\alpha$-rich stars.
    \emph{Orange}: old and metal-rich.
    \emph{Red}: secondary RC stars.
    \emph{Magenta}: Group Four (KIC\,6605673 not shown).
    The values for our sample are those adopted in this work (see Tables~\ref{tab:tab1} and \ref{tab:atmparam}).
    }
    \label{fig:sample}
\end{figure}

The stars present in this work are solar-like oscillators with available high-quality \emph{Kepler} photometry (see Table~\ref{tab:tab1} for their respective $\Delta \nu$ and $\nu_{\mathrm{max}}$ values) and [M/H]\footnote{The square bracket notation [A/B] represents the logarithmic abundance ratio between species A and B normalised by the solar value: [A/B] = log$_{10}$(A/B) - log$_{10}$(A/B)$_{\odot}$.} $>$ -0.5 dex -- the more metal-poor objects from the observing run were analysed in Paper~I. One of them -- KIC\,6605673 -- is a main-sequence turnoff object, while the rest are red giants. The former was observed by \emph{Kepler} in high-cadence mode, while the evolved stars have observations in low-cadence mode. All targets except KIC\,6634419 have parallax measurements in both Gaia DR2 and (E)DR3 \citep{GaiaDR2:2018,GaiaEDR3:2021,2022arXiv220800211G}. KIC\,6634419 has a parallax measurement only in DR3, with precision of 17\%, while DR2 precision in parallaxes are always below 7\% in our sample. Fig.~\ref{fig:sample} shows how they compare to other giants in data from asteroseismic surveys.

The stars included in this study are divided into four groups: (i) \emph{APOKASC young $\alpha$-rich stars}, red giants previously classified as young and $\alpha$-rich by \citet[][]{Martig:2015aa}, and discussed in detail in Section~\ref{sec:youngalpharich}, (ii) other group of red giants identified as \emph{old and metal-rich} by \citet[][]{Casagrande:2014aa}, analysed in Section~\ref{sec:omr}, as well as (iii) four secondary clump objects with strong mode damping as evident by the very broad peaks in the frequency power spectra of their Kepler light curves, referred here as \emph{secondary RC stars}. The last (iv) group (\emph{Group Four}) contains three stars that do not fit in any of the previous three groups, and are discussed in Section~\ref{sec:other} together with the secondary RC stars group. One of the Group Four stars, KIC\,6940126, had been previously identified as old and metal-rich by \citet[][]{Casagrande:2014aa}, but discussed separately in Section~\ref{sec:resultages} due to its age uncertainty and location of its power excess near the Nyquist frequency. The other two were also studied by \citeauthor{Casagrande:2014aa}, KIC\,8145677 being identified as old and metal-poor, and KIC\,6605673, classified as a metal-poor dwarf in their work.

\section{Analysis}
\label{sec:analysis}

This work combines chemical analysis from spectroscopic observations, grid-based asteroseismic modelling using \emph{Kepler} data, and kinematics from Gaia. The methods are summarised in the following paragraphs. We refer to Paper~I for a full methodological description, as both works share the same methodology.

Chemical abundances for 26 elements (Li, O, Na, Mg, Al, Si, Ca, Sc, Ti, V, Cr, Mn, Fe, Co, Ni, Cu, Zn, Sr, Y, Zr, Ba, La, Ce, Nd, Sm, Eu) were measured through the classical spectroscopic method, employing Gaussian fitting on weak isolated spectral lines for measuring their equivalent widths, whose values are shown in the supplementary material. For lines with blending and/or hyperfine splitting, spectral synthesis was employed -- i.e., a synthetic spectrum was fitted using the 2017 version of \textsc{moog} \citep{Sneden:1973phd} to the observed one for each line under consideration. The line list was compiled from several sources \citep[][]{Cayrel:2004aa,Alves-Brito:2010aa,Melendez:2012,Barbuy:2013,Ishigaki:2013,Yong:2014}, and the log~gf values from these sources were compared with those published in the NIST database\footnote{\url{https://physics.nist.gov/PhysRefData/ASD/lines_form.html}}. The log~gf values from NIST were adopted when there was disagreement.

The atmospheric models required to translate equivalent widths in chemical abundances, as well as to build synthetic spectra, were interpolated from a solar-scaled plane-parallel 1D LTE ATLAS9/ODFNEW grid without overshooting \citep[][]{Castelli:2003aa}. The input atmospheric parameters required for interpolation (effective temperature T$_{\mathrm{eff}}$, surface gravity log~g, metallicity [M/H] and microturbulent velocity $v_t$) were defined as follows. First, a set of 'pure' spectroscopic atmospheric parameters was calculated with \textsc{moog} using \ion{Fe}{I} and \ion{Fe}{II} excitation/ionisation balance. [M/H] was estimated from [Fe/H] and [$\alpha$/Fe] using the formula from \citet{Salaris:1993}. The estimated log~g and [M/H], as well as reddening from the 3D map from \citet[][]{Green:2019} and distances from \citet[][]{BailerJones:2018}, were used to feed the InfraRed Flux Method \citep[IRFM, ][]{Blackwell:1977aa,Casagrande:2010aa,Casagrande:2021} employed to derive T$_{\mathrm{eff}}$. Then, with [M/H] and the updated T$_{\mathrm{eff}}$, log~g was updated using asteroseismic information (see discussion in the next paragraph). New values of [M/H] and $v_t$ were calculated with \textsc{moog}, this time keeping the IRFM T$_{\mathrm{eff}}$ and the asteroseismic log~g fixed. A new iteration was performed for each star until convergence was reached. Uncertainties for [M/H] and $v_t$ were calculated adapting the method from \citet[][see Paper~I for details]{Epstein:2010aa}, while T$_{\mathrm{eff}}$ uncertainties were propagated from the IRFM inputs using MonteCarlo. The adopted atmospheric parameters are shown in Table~\ref{tab:atmparam}. When compared to the initial spectroscopic guess, the IRFM effective temperatures tend to be slightly cooler (median difference is -71~$\pm$~54~K), and the asteroseismic log~g show a median difference of -0.12~$\pm$~0.27~dex with respect to the spectroscopic ones. The median adopted values for metallicity and microturbulence differ by a small amount (-0.04~dex and -0.02 km s$^{-1}$, respectively), and present a small scatter as well (0.05~dex and 0.04 km s$^{-1}$). No trend between the differences in the methods with respect to the atmospheric parameters was detected, and the Pearson correlation coefficient between the results from the different methods is greater than 0.9 in all the four cases.

\begin{table*}
	\centering
	\caption{
	The stellar parameters adopted to generate the atmospheric models for the spectroscopic analysis and their respective uncertainties.}
	\label{tab:atmparam}
	\begin{tabular}{lrrrr}
		\hline
		Star & T$_{\mathrm{eff}}$ & log g & [M/H] & $v_t$ \\
             & K & dex & dex & km s$^{-1}$ \\
        \hline
             KIC\,2845610 & 5236\;$\pm$\;\;98 & 2.851 $\pm$ 0.006 &  0.05 $\pm$ 0.06 & 1.29 $\pm$ 0.05 \\
             KIC\,3455760 & 4728\;$\pm$109    & 2.589 $\pm$ 0.004 &  0.14 $\pm$ 0.08 & 1.19 $\pm$ 0.09 \\
             KIC\,3833399 & 4781\;$\pm$\;\;68 & 2.472 $\pm$ 0.008 &  0.16 $\pm$ 0.06 & 1.45 $\pm$ 0.14 \\
             KIC\,5512910 & 4911\;$\pm$\;\;95 & 2.491 $\pm$ 0.009 & -0.19 $\pm$ 0.06 & 1.42 $\pm$ 0.05 \\
             KIC\,5707338 & 5057\;$\pm$\;\;85 & 2.812 $\pm$ 0.003 &  0.23 $\pm$ 0.05 & 1.30 $\pm$ 0.10 \\
             KIC\,6605673 & 5990\;$\pm$\;\;58 & 4.050 $\pm$ 0.008 & -0.06 $\pm$ 0.03 & 1.05 $\pm$ 0.10 \\
             KIC\,6634419 & 5335\;$\pm$135    & 2.867 $\pm$ 0.004 &  0.19 $\pm$ 0.11 & 1.02 $\pm$ 0.18 \\
             KIC\,6936796 & 4521\;$\pm$\;\;60 & 2.245 $\pm$ 0.012 &  0.19 $\pm$ 0.06 & 1.89 $\pm$ 0.15 \\
             KIC\,6940126 & 4831\;$\pm$\;\;52 & 3.316 $\pm$ 0.003 &  0.26 $\pm$ 0.05 & 1.15 $\pm$ 0.14 \\
             KIC\,7595155 & 4568\;$\pm$\;\;58 & 2.357 $\pm$ 0.009 &  0.28 $\pm$ 0.07 & 1.66 $\pm$ 0.21 \\
             KIC\,8145677 & 5112\;$\pm$\;\;66 & 2.406 $\pm$ 0.005 & -0.41 $\pm$ 0.04 & 1.56 $\pm$ 0.05 \\
             KIC\,9002884 & 4218\;$\pm$\;\;50 & 1.549 $\pm$ 0.002 & -0.14 $\pm$ 0.03 & 1.60 $\pm$ 0.12 \\
             KIC\,9266192 & 5119\;$\pm$\;\;69 & 2.795 $\pm$ 0.006 &  0.14 $\pm$ 0.05 & 1.36 $\pm$ 0.11 \\
             KIC\,9761625 & 4413\;$\pm$\;\;49 & 1.853 $\pm$ 0.014 & -0.02 $\pm$ 0.03 & 1.45 $\pm$ 0.10 \\
            KIC\,10525475 & 4845\;$\pm$\;\;59 & 2.494 $\pm$ 0.012 &  0.01 $\pm$ 0.04 & 1.27 $\pm$ 0.10 \\
            KIC\,11823838 & 4910\;$\pm$\;\;72 & 2.539 $\pm$ 0.008 & -0.18 $\pm$ 0.04 & 1.41 $\pm$ 0.05 \\
	    \hline
	\end{tabular}
\end{table*}

Stellar ages, masses, radii, as well as log~g, were derived with \textsc{basta} \citep[BAyesian STellar Algorithm,][]{SilvaAguirre:2015,SilvaAguirre:2017,Aguirre:2022}. A set of observables were fitted for each star using the grid of BaSTI isochrones from \citep{Hidalgo:2018} and their mass loss prescription $\eta =$~0.3. The chosen observables are equivalent to the s$_n$ set from Paper~I, i.e., the average asteroseismic parameters $\Delta \nu$ and $\nu_{\mathrm{max}}$ \citep[with the correction from][]{Serenelli:2017}, T$_{\mathrm{eff}}$, [M/H], photometry in the 2MASS K$_s$ band \citep[][]{Cutri:2003}, and a prior on the evolutionary phase which, for red giants, depends on the object position in the period spacing versus $\Delta \nu$ diagram \citep[following][]{Mosser:2014} to differentiate those ascending the Red Giant Branch (RGB) from those in the Core-Helium Burning phase (CHeB). The uncertainties for the \textsc{basta}-derived parameters were taken from the 16th- and 84th- percentiles of their posterior distributions.

The sources of the asteroseismic parameters $\Delta \nu$ and $\nu_{\mathrm{max}}$ are listed in Table~\ref{tab:tab1}. For KIC\,6940126 $\Delta \nu$ and $\nu_{\mathrm{max}}$ were estimated following the approach of \citet[][]{Lund:2016}, adopting for $\nu_{\mathrm{max}}$ the fit of background model given by a sum of generalised Lorentzian functions with free exponents \citep[][]{Harvey:1985,Karoff:2012,Lund:2017}, including a Gaussian envelope centred on $\nu_{\mathrm{max}}$ to take into account the power excess from oscillations. The $\Delta \nu$ was measured from the peak of the power-of-power spectrum ($\rm PS \otimes PS$) centred on $\Delta \nu$/2, and the full width at half maximum of that peak corresponds to the $\Delta \nu$ uncertainty.

As in Paper~I, orbital properties were determined for the sample using the \textsc{Python}-based galactic dynamics package, \textsc{GalPy} \citep[][]{Bovy:2015}. Orbits were initialised for each star using the astrometric properties from Gaia DR3, with the zero-point correction for parallaxes from \citet[][]{Lindegren:2021}. Each orbit was integrated forward and backwards for 2~Gyr in the static Galactic potential of \citet[][]{McMillan:2017}. Galactic coordinates, local standard of rest and Solar position remained the same as in Paper~I \citep[][]{Schonrich:2010}. In total, each orbit was integrated for 10-14 Gyr to determine the pericentric radii, apocentric radii, orbital energy, orbital eccentricity and maximum height from the plane (|Z|$_{\mathrm{max}}$). The radial (J$_r$), azimuthal (equivalent to the z-component of the angular momentum, L$_{\mathrm{z}}$) and z-actions (J$_z$) were determined using the Staekel fudge \citep[][]{Binney:2012,Mackereth:2018}. Finally, uncertainties were determined for each orbital property through determination of the covariance matrix associated with the Gaia DR3 astrometry. To sample the full error distribution of each input parameter, 1000 MonteCarlo realisations were made assuming a symmetric error distribution for the radial velocities.

\section{Results and discussion}
\label{sec:results}

\begin{figure*}
	\includegraphics{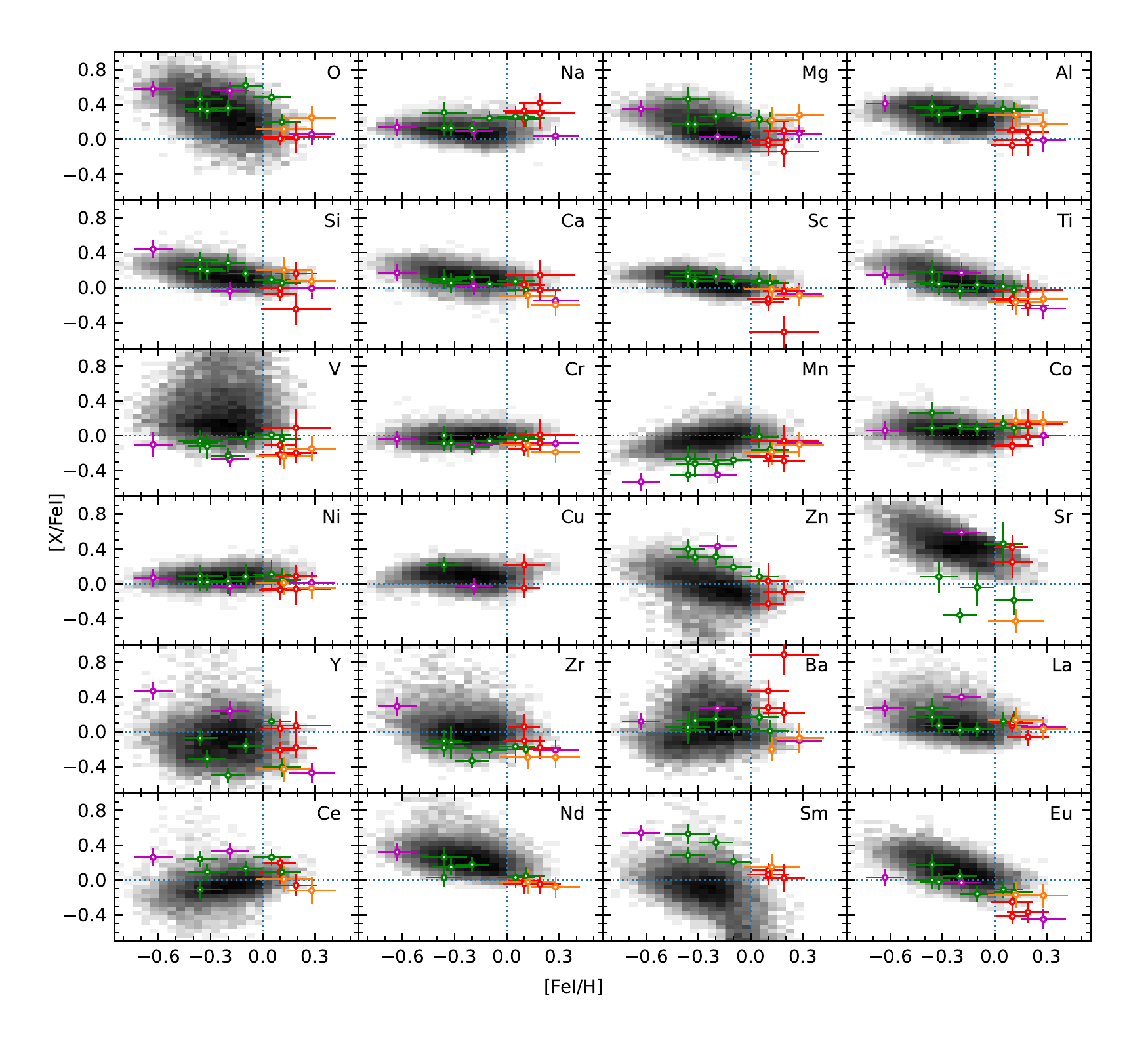}
    \caption{Abundance ratios [X/\ion{Fe}{I}] as function of [\ion{Fe}{I}/H] for all elements measured in this work other than Li.
    Point colours follow Fig.~\ref{fig:sample}.
    The greyscale histograms represent the density of $\sim$ 9000 GALAH~DR3 red giants \citep{Buder:2021} in the [X/Fe]-[Fe/H] space, shown here for guidance. Histogram intensity scales to log(N), where N is the number of stars from the selected GALAH sample. Blue dotted lines represent solar abundances. Histogram bin size is 0.05~$\times$~0.05~dex.
    }
    \label{fig:abundances}
\end{figure*}

\begin{table*}
	\centering
	\caption{Abundance ratios A(Li) and [X/H]. Solar abundances are from \citet{Asplund:2009aa}.}
\label{tab:abundances1}
\begin{tabular}{rrrrrrrrrr}
\hline
     KIC &            A(Li) &            [O/H] &           [Na/H] &           [Mg/H] &           [Al/H] &           [Si/H] &           [Ca/H] &           [Sc/H] &           [Ti/H] \\
\hline 
 2845610 &         $\cdots$ &  0.11 $\pm$ 0.05 &  0.33 $\pm$ 0.09 &  0.04 $\pm$ 0.14 &  0.21 $\pm$ 0.14 &  0.02 $\pm$ 0.06 &  0.14 $\pm$ 0.08 & -0.07 $\pm$ 0.04 & -0.05 $\pm$ 0.12 \\ 
 3455760 &         $\cdots$ &  0.53 $\pm$ 0.05 &  0.31 $\pm$ 0.14 &  0.28 $\pm$ 0.10 &  0.40 $\pm$ 0.09 &  0.14 $\pm$ 0.12 &  0.12 $\pm$ 0.19 &  0.13 $\pm$ 0.04 &  0.06 $\pm$ 0.16 \\ 
 3833399 &         $\cdots$ &  0.31 $\pm$ 0.03 &  0.36 $\pm$ 0.09 &  0.32 $\pm$ 0.11 &  0.44 $\pm$ 0.06 &  0.16 $\pm$ 0.11 &  0.08 $\pm$ 0.08 &  0.16 $\pm$ 0.08 &  0.08 $\pm$ 0.12 \\ 
 5512910 &         $\cdots$ &  0.01 $\pm$ 0.05 & -0.19 $\pm$ 0.07 & -0.15 $\pm$ 0.04 & -0.04 $\pm$ 0.07 & -0.13 $\pm$ 0.10 & -0.27 $\pm$ 0.12 & -0.24 $\pm$ 0.04 & -0.28 $\pm$ 0.14 \\ 
 5707338 &         $\cdots$ &  0.21 $\pm$ 0.02 &  0.61 $\pm$ 0.12 &  0.29 $\pm$ 0.11 &  0.27 $\pm$ 0.06 &  0.35 $\pm$ 0.13 &  0.16 $\pm$ 0.10 &  0.15 $\pm$ 0.07 & -0.02 $\pm$ 0.11 \\ 
 6605673 &  2.61 $\pm$ 0.07 &  0.37 $\pm$ 0.07 & -0.09 $\pm$ 0.12 & -0.16 $\pm$ 0.05 &   $\cdots$ & -0.23 $\pm$ 0.03 & -0.17 $\pm$ 0.10 & -0.07 $\pm$ 0.05 & -0.02 $\pm$ 0.12 \\ 
 6634419 &         $\cdots$ &  0.21 $\pm$ 0.06 &  0.49 $\pm$ 0.12 &  0.05 $\pm$ 0.05 &  0.18 $\pm$ 0.12 & -0.06 $\pm$ 0.03 &  0.33 $\pm$ 0.13 & -0.32 $\pm$ 0.05 &  0.16 $\pm$ 0.15 \\ 
 6936796 &         $\cdots$ &  0.24 $\pm$ 0.02 &   $\cdots$ &  0.34 $\pm$ 0.01 &  0.40 $\pm$ 0.09 &  0.32 $\pm$ 0.13 &  0.03 $\pm$ 0.11 &  0.10 $\pm$ 0.13 & -0.05 $\pm$ 0.11 \\ 
 6940126 &         $\cdots$ &  0.34 $\pm$ 0.01 &  0.32 $\pm$ 0.07 &  0.35 $\pm$ 0.05 &  0.27 $\pm$ 0.13 &  0.27 $\pm$ 0.11 &  0.13 $\pm$ 0.09 &  0.21 $\pm$ 0.06 &  0.04 $\pm$ 0.10 \\ 
 7595155 &  0.35 $\pm$ 0.11 &  0.53 $\pm$ 0.02 &   $\cdots$ &  0.56 $\pm$ 0.07 &  0.45 $\pm$ 0.14 &  0.35 $\pm$ 0.13 &  0.08 $\pm$ 0.08 &  0.19 $\pm$ 0.09 &  0.15 $\pm$ 0.12 \\ 
 8145677 &         $\cdots$ & -0.05 $\pm$ 0.06 & -0.49 $\pm$ 0.05 & -0.28 $\pm$ 0.06 & -0.22 $\pm$ 0.08 & -0.19 $\pm$ 0.09 & -0.46 $\pm$ 0.07 &   $\cdots$ & -0.49 $\pm$ 0.10 \\ 
 9002884 &         $\cdots$ &  0.10 $\pm$ 0.03 & -0.05 $\pm$ 0.08 &  0.10 $\pm$ 0.15 &  0.02 $\pm$ 0.12 & -0.15 $\pm$ 0.12 & -0.26 $\pm$ 0.09 & -0.23 $\pm$ 0.03 & -0.18 $\pm$ 0.14 \\ 
 9266192 &         $\cdots$ &   $\cdots$ &  0.43 $\pm$ 0.08 &  0.09 $\pm$ 0.12 &  0.03 $\pm$ 0.13 &  0.09 $\pm$ 0.06 &  0.13 $\pm$ 0.10 & -0.03 $\pm$ 0.15 & -0.03 $\pm$ 0.08 \\ 
 9761625 &         $\cdots$ &  0.16 $\pm$ 0.02 & -0.07 $\pm$ 0.05 &  0.06 $\pm$ 0.04 &  0.11 $\pm$ 0.04 &  0.08 $\pm$ 0.11 & -0.08 $\pm$ 0.09 & -0.07 $\pm$ 0.08 & -0.24 $\pm$ 0.08 \\ 
10525475 &         $\cdots$ &  0.52 $\pm$ 0.01 &  0.14 $\pm$ 0.05 &  0.18 $\pm$ 0.11 &  0.23 $\pm$ 0.06 &  0.06 $\pm$ 0.07 & -0.06 $\pm$ 0.10 & -0.03 $\pm$ 0.08 & -0.07 $\pm$ 0.12 \\ 
11823838 &         $\cdots$ & -0.01 $\pm$ 0.07 & -0.23 $\pm$ 0.05 & -0.18 $\pm$ 0.13 &  0.00 $\pm$ 0.09 & -0.04 $\pm$ 0.08 & -0.27 $\pm$ 0.09 & -0.19 $\pm$ 0.04 & -0.30 $\pm$ 0.09 \\ 
\hline
     KIC &            [V/H] &           [Cr/H] &           [Mn/H] &          [FeI/H] &         [FeII/H] &           [Co/H] &           [Ni/H] &           [Cu/H] &           [Zn/H] \\
\hline
 2845610 & -0.01 $\pm$ 0.14 & -0.05 $\pm$ 0.08 & -0.15 $\pm$ 0.06 &  0.10 $\pm$ 0.09 &  0.01 $\pm$ 0.10 & -0.01 $\pm$ 0.10 &  0.12 $\pm$ 0.10 &  0.05 $\pm$ 0.13 & -0.13 $\pm$ 0.05 \\
 3455760 &  0.06 $\pm$ 0.18 &  0.03 $\pm$ 0.11 &  0.04 $\pm$ 0.15 &  0.05 $\pm$ 0.11 &  0.00 $\pm$ 0.18 &  0.19 $\pm$ 0.05 &  0.16 $\pm$ 0.19 &   $\cdots$ &  0.13 $\pm$ 0.09 \\
 3833399 &  0.07 $\pm$ 0.12 &  0.07 $\pm$ 0.08 & -0.05 $\pm$ 0.09 &  0.11 $\pm$ 0.11 &  0.06 $\pm$ 0.10 &  0.20 $\pm$ 0.06 &  0.15 $\pm$ 0.16 &   $\cdots$ &   $\cdots$ \\
 5512910 & -0.44 $\pm$ 0.16 & -0.40 $\pm$ 0.10 & -0.64 $\pm$ 0.18 & -0.32 $\pm$ 0.11 & -0.24 $\pm$ 0.12 &   $\cdots$ & -0.30 $\pm$ 0.08 &   $\cdots$ & -0.02 $\pm$ 0.08 \\
 5707338 & -0.01 $\pm$ 0.12 &  0.11 $\pm$ 0.07 & -0.10 $\pm$ 0.14 &  0.19 $\pm$ 0.12 &  0.20 $\pm$ 0.17 &  0.17 $\pm$ 0.07 &  0.28 $\pm$ 0.13 &   $\cdots$ &  0.10 $\pm$ 0.09 \\
 6605673 & -0.46 $\pm$ 0.05 & -0.30 $\pm$ 0.04 & -0.64 $\pm$ 0.07 & -0.19 $\pm$ 0.11 &  0.10 $\pm$ 0.08 &   $\cdots$ & -0.22 $\pm$ 0.10 & -0.22 $\pm$ 0.04 &  0.24 $\pm$ 0.13 \\
 6634419 &  0.28 $\pm$ 0.22 &  0.20 $\pm$ 0.14 &  0.13 $\pm$ 0.17 &  0.19 $\pm$ 0.20 &  0.33 $\pm$ 0.17 &  0.32 $\pm$ 0.13 &  0.13 $\pm$ 0.15 &   $\cdots$ &   $\cdots$ \\
 6936796 & -0.12 $\pm$ 0.09 &  0.01 $\pm$ 0.09 & -0.07 $\pm$ 0.11 &  0.12 $\pm$ 0.16 &  0.12 $\pm$ 0.15 &  0.29 $\pm$ 0.09 &  0.13 $\pm$ 0.09 &   $\cdots$ &   $\cdots$ \\
 6940126 &  0.13 $\pm$ 0.07 &  0.19 $\pm$ 0.08 &  0.19 $\pm$ 0.14 &  0.28 $\pm$ 0.13 &  0.24 $\pm$ 0.13 &  0.28 $\pm$ 0.06 &  0.29 $\pm$ 0.13 &   $\cdots$ &   $\cdots$ \\
 7595155 &  0.13 $\pm$ 0.14 &  0.09 $\pm$ 0.07 &  0.18 $\pm$ 0.15 &  0.28 $\pm$ 0.14 &  0.14 $\pm$ 0.12 &  0.44 $\pm$ 0.05 &  0.23 $\pm$ 0.10 &   $\cdots$ &   $\cdots$ \\
 8145677 & -0.73 $\pm$ 0.16 & -0.67 $\pm$ 0.05 & -1.16 $\pm$ 0.09 & -0.63 $\pm$ 0.11 & -0.69 $\pm$ 0.12 & -0.57 $\pm$ 0.10 & -0.56 $\pm$ 0.09 &   $\cdots$ &   $\cdots$ \\
 9002884 & -0.42 $\pm$ 0.12 & -0.36 $\pm$ 0.09 & -0.63 $\pm$ 0.16 & -0.36 $\pm$ 0.13 & -0.27 $\pm$ 0.19 & -0.10 $\pm$ 0.01 & -0.27 $\pm$ 0.12 &   $\cdots$ &   $\cdots$ \\
 9266192 & -0.12 $\pm$ 0.10 &  0.02 $\pm$ 0.07 & -0.14 $\pm$ 0.13 &  0.10 $\pm$ 0.12 &  0.22 $\pm$ 0.11 & -0.02 $\pm$ 0.11 &  0.03 $\pm$ 0.12 &  0.32 $\pm$ 0.12 &  0.13 $\pm$ 0.24 \\
 9761625 & -0.43 $\pm$ 0.06 & -0.33 $\pm$ 0.06 & -0.52 $\pm$ 0.12 & -0.20 $\pm$ 0.10 & -0.08 $\pm$ 0.17 & -0.09 $\pm$ 0.02 & -0.17 $\pm$ 0.17 &   $\cdots$ &  0.11 $\pm$ 0.08 \\
10525475 & -0.14 $\pm$ 0.11 & -0.16 $\pm$ 0.06 & -0.38 $\pm$ 0.08 & -0.10 $\pm$ 0.10 & -0.13 $\pm$ 0.13 & -0.02 $\pm$ 0.04 & -0.02 $\pm$ 0.16 &   $\cdots$ &  0.09 $\pm$ 0.08 \\
11823838 & -0.45 $\pm$ 0.13 & -0.45 $\pm$ 0.06 & -0.81 $\pm$ 0.07 & -0.36 $\pm$ 0.10 & -0.25 $\pm$ 0.14 & -0.27 $\pm$ 0.08 & -0.34 $\pm$ 0.11 & -0.14 $\pm$ 0.06 &  0.04 $\pm$ 0.13 \\
\hline
     KIC &           [Sr/H] &            [Y/H] &           [Zr/H] &           [Ba/H] &           [La/H] &           [Ce/H] &           [Nd/H] &           [Sm/H] &           [Eu/H] \\
\hline
 2845610 &  0.52 $\pm$ 0.16 & -0.11 $\pm$ 0.11 &  0.16 $\pm$ 0.17 &  0.38 $\pm$ 0.10 &  0.17 $\pm$ 0.05 &  0.30 $\pm$ 0.06 &  0.09 $\pm$ 0.08 &  0.21 $\pm$ 0.08 & -0.32 $\pm$ 0.08 \\
 3455760 &  0.51 $\pm$ 0.29 &  0.17 $\pm$ 0.06 & -0.12 $\pm$ 0.22 &  0.22 $\pm$ 0.10 &  0.17 $\pm$ 0.10 &  0.31 $\pm$ 0.07 &  0.08 $\pm$ 0.13 &   $\cdots$ & -0.06 $\pm$ 0.03 \\
 3833399 & -0.08 $\pm$ 0.19 & -0.30 $\pm$ 0.11 & -0.09 $\pm$ 0.13 &  0.12 $\pm$ 0.14 &  0.25 $\pm$ 0.08 &  0.20 $\pm$ 0.09 &  0.16 $\pm$ 0.09 &   $\cdots$ & -0.03 $\pm$ 0.01 \\
 5512910 & -0.24 $\pm$ 0.21 & -0.63 $\pm$ 0.03 & -0.44 $\pm$ 0.22 & -0.19 $\pm$ 0.05 & -0.25 $\pm$ 0.07 & -0.23 $\pm$ 0.06 & -0.17 $\pm$ 0.07 &   $\cdots$ & -0.34 $\pm$ 0.02 \\
 5707338 &   $\cdots$ &  0.01 $\pm$ 0.11 &  0.01 $\pm$ 0.15 &  0.41 $\pm$ 0.12 &  0.13 $\pm$ 0.05 &  0.13 $\pm$ 0.14 &  0.14 $\pm$ 0.08 &  0.21 $\pm$ 0.18 & -0.18 $\pm$ 0.00 \\
 6605673 &  0.40 $\pm$ 0.05 &  0.05 $\pm$ 0.11 &   $\cdots$ &  0.08 $\pm$ 0.04 &  0.21 $\pm$ 0.11 &  0.14 $\pm$ 0.03 &   $\cdots$ &   $\cdots$ & -0.22 $\pm$ 0.04 \\
 6634419 &   $\cdots$ &  0.26 $\pm$ 0.12 &   $\cdots$ &  1.08 $\pm$ 0.25 &   $\cdots$ &   $\cdots$ &   $\cdots$ &   $\cdots$ &   $\cdots$ \\
 6936796 & -0.31 $\pm$ 0.10 & -0.31 $\pm$ 0.10 & -0.17 $\pm$ 0.10 & -0.08 $\pm$ 0.10 &  0.26 $\pm$ 0.05 &  0.13 $\pm$ 0.14 &  0.10 $\pm$ 0.09 &  0.27 $\pm$ 0.05 & -0.05 $\pm$ 0.02 \\
 6940126 &   $\cdots$ & -0.19 $\pm$ 0.10 &  0.07 $\pm$ 0.08 &  0.18 $\pm$ 0.14 &  0.34 $\pm$ 0.03 &   $\cdots$ &   $\cdots$ &   $\cdots$ & -0.17 $\pm$ 0.04 \\
 7595155 &   $\cdots$ &   $\cdots$ & -0.01 $\pm$ 0.08 &  0.21 $\pm$ 0.19 &  0.31 $\pm$ 0.06 &  0.16 $\pm$ 0.17 &  0.20 $\pm$ 0.08 &   $\cdots$ &  0.10 $\pm$ 0.01 \\
 8145677 &   $\cdots$ & -0.16 $\pm$ 0.09 & -0.34 $\pm$ 0.11 & -0.51 $\pm$ 0.06 & -0.36 $\pm$ 0.05 & -0.37 $\pm$ 0.02 & -0.31 $\pm$ 0.09 & -0.09 $\pm$ 0.07 & -0.60 $\pm$ 0.04 \\
 9002884 &   $\cdots$ &   $\cdots$ & -0.54 $\pm$ 0.06 & -0.31 $\pm$ 0.22 & -0.19 $\pm$ 0.05 & -0.47 $\pm$ 0.10 & -0.10 $\pm$ 0.08 &  0.17 $\pm$ 0.10 & -0.18 $\pm$ 0.04 \\
 9266192 &  0.35 $\pm$ 0.21 &  0.14 $\pm$ 0.09 &  0.00 $\pm$ 0.13 &  0.57 $\pm$ 0.13 &  0.20 $\pm$ 0.04 &  0.19 $\pm$ 0.14 &  0.06 $\pm$ 0.13 &  0.16 $\pm$ 0.02 & -0.15 $\pm$ 0.06 \\
 9761625 & -0.56 $\pm$ 0.07 & -0.70 $\pm$ 0.06 & -0.53 $\pm$ 0.08 & -0.05 $\pm$ 0.16 & -0.18 $\pm$ 0.03 &   $\cdots$ & -0.02 $\pm$ 0.04 &  0.23 $\pm$ 0.04 & -0.16 $\pm$ 0.02 \\
10525475 & -0.14 $\pm$ 0.24 & -0.26 $\pm$ 0.06 & -0.31 $\pm$ 0.10 & -0.07 $\pm$ 0.12 & -0.07 $\pm$ 0.07 &  0.03 $\pm$ 0.06 &   $\cdots$ &  0.11 $\pm$ 0.06 & -0.26 $\pm$ 0.03 \\
11823838 &   $\cdots$ & -0.43 $\pm$ 0.02 & -0.46 $\pm$ 0.13 & -0.34 $\pm$ 0.05 & -0.09 $\pm$ 0.14 & -0.12 $\pm$ 0.03 & -0.33 $\pm$ 0.11 & -0.08 $\pm$ 0.09 & -0.36 $\pm$ 0.01 \\
\hline
\end{tabular}

\end{table*}

Fig.~\ref{fig:abundances} shows all abundances in terms of [X/Fe] derived in this work plotted as function of [Fe/H] -- reference solar values are from \citet[][]{Asplund:2009aa}. For comparison, abundance ratios of $\sim$9000 red giants from GALAH \citep{Buder:2021}, selected in the same range of stellar parameters and parallaxes of the targets in our sample, are shown as greyscale clouds. For the majority of elements, the results from this study usually fall on the same range of abundance ratios for the respective metallicites in GALAH. For a handful of stars, there are possible outliers which lie on the edge of the GALAH distribution for the elements O, Na and Al. For the elements V, Mn, Zn and Sm, most of the sample appear offset from the GALAH distribution. Finally, for Sr and Y the sample exhibits a large scatter when compared to other elements.

The O abundances were obtained from the [\ion{O}{I}] lines at 6300~\AA~and 6363~\AA, measured by spectral synthesis. The 6300-line has a known blend with a Ni feature, and, for that reason, the spectral synthesis took into account the Ni abundances previously measured though EW. As seen in Fig.~\ref{fig:abundances}, the objects with anomalous O abundances do not have corresponding Ni anomalies, thus systematics arising from line blending could be discarded in their cases.

As for Na and Al, their abundances come from the usual lines employed in the literature for red giants: the \ion{Na}{I} lines\footnote{For the main-sequence star KIC\,6605673 and the most metal-poor giant in the sample (KIC\,8145677), measurements of the \ion{Na}{I} lines at 5682~\AA~and 5688~\AA~are included as well.} at 6154~\AA~and 6160~\AA, as well as the \ion{Al}{I} features at 6696~\AA~and 6698~\AA. Although there is an average enhancement at solar metallicity -- $\sim$~0.2~dex in Na and $\sim$~0.3~dex in Al, most data points fit inside the GALAH clouds in Fig.~\ref{fig:abundances} for these species. Meanwhile, the distribution of Mn abundances -- measured from the 6100~\AA~triplet with their hyperfine structure calculated using the same constants adopted by \citet[][]{Barbuy:2013} -- follow the evolution of [Mn/Fe] versus [Fe/H] of the GALAH cloud, but in most cases shifted by at least $-$0.2~dex. The likely cause of that offset in Mn is the choice of log~gf values, taken from the NIST database as mentioned in Section~\ref{sec:analysis}, which may be overestimated. This negative shift is apparent in the 6645~\AA~\ion{Eu}{II} abundances as well, albeit to a lesser extent. The results for [Mg/Fe] follow the typical distinction between 'high' and 'normal' Mg, which is not clear in the other individual $\alpha$ elements, Ca in particular. Two of the APOKASC young $\alpha$-rich giants -- KIC\,5512910, and KIC\,11823838, and KIC\,6605673, from Group Four, seem to fit clearly the lower (normal) envelope of the [Mg/Fe] GALAH distribution.

\begin{figure}
	\includegraphics{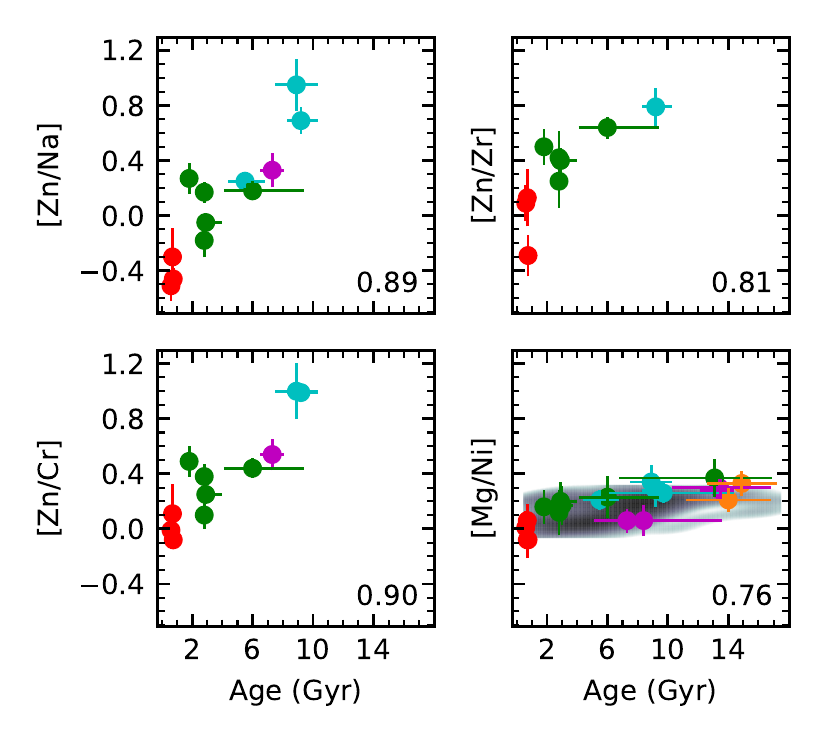}
    \caption{Selected abundance ratios which correlate with age.
    Numbers in the bottom right of each plot are the corresponding Pearson correlation coefficient. In all cases shown here the corresponding p-value is lower than 0.01.
    The density map in the bottom right plot combines APOKASC-2 and APOGEE-2 data \citep[][]{Pinsonneault:2018,Jonsson:2020}, and their colour map scales to log(N), where N is the number of stars -- darker representing higher density of objects.
    Colours as in Fig.~\ref{fig:sample}, cyan circles are stars from Paper~I. Ages for our data points were derived from pure asteroseismic solutions (see Section~\ref{sec:resultages}).
    }
    \label{fig:zn_corr}
\end{figure}

An element with intriguing results is Zn, despite the apparent $\sim$~0.2~dex zero-point shift seen in Fig.~\ref{fig:abundances} for [Zn/Fe]. In our dataset several abundance ratios that include Zn correlate with stellar ages -- the strongest of them (Na, Cr, Zr) are shown in Fig.~\ref{fig:zn_corr} with their respective correlation coefficients. These Zn-abundance ratios display steep gradients for chemical clocks, $\approx$ 0.1 dex Gyr$^{-1}$ from least-squares linear fits. Among the stars with Zn measurements, the metal-poor ones KIC\,11563791 and KIC\,4345370 (both analysed in Paper~I) display strong enhancement in this species. Even if this pair is removed the correlation with age persists -- Pearson correlation coefficient $\approx$~0.75 in the three cases (when the two metal-poor stars from Paper~I are ignored), p-value $<$~0.01 in [Zn/Na], and $<$~0.03 in [Zn/Cr] and [Zn/Zr]. Since our sample is relatively small and with strong selection effects, these intriguing correlations could be spurious.

In order to put the Zn results in perspective, the bottom right plot of Fig.~\ref{fig:zn_corr} shows [Mg/Ni] as function of age. Despite also showing a strong correlation with age (Pearson coefficient is 0.76), the young and $\alpha$-rich stars (green markers), as well as the metal-poor targets studied in Paper~I (cyan markers), show enhancement in [Mg/Ni] when compared to APOGEE-2/APOKASC-2 results \citep{Jonsson:2020,Pinsonneault:2018}. This may suggest that the stars in this study would not be suitable as chemical clocks, as, when we inspect our results against the APOGEE-2/APOKASC-2 results, our sample seems to be divided in high-[Mg/Ni] and normal-[Mg/Ni], the 'normal' ones matching the space of highest density of the comparison sample ([Mg/Ni] $\lesssim$ +0.10). Nevertheless, a comprehensive high-precision study of Zn abundances in red giants with robust age determination \citep[as done for solar twins by, e.g.,][]{Spina:2016aa} could be extremely valuable.

We did check the magnitude of NLTE departures in four representative stars -- KIC\,2845610, KIC\,6936796, KIC\,8145677, KIC\,9002884 -- for the species available in the INSPECT database\footnote{\url{http://www.inspect-stars.com/}, version 1.0.}. We did find lines in common with our line list for Na, Mg, Ti, \ion{Fe}{I}, and \ion{Fe}{II} \citep[][]{Bergemann:2011aa,Lind:2011,Lind:2012aa,Bergemann:2012aa,Osorio:2015,Osorio:2016}. The departures are mild except for Na, ranging between zero and $\sim$+0.05~dex. Thus, for most of the [X/\ion{Fe}{I}] abundance ratios the NLTE departures of the individual species tend to cancel out. \ion{Fe}{II} NLTE corrections are estimated as -0.01~dex for all stars tested. For A(Na), the NLTE departures range from $\sim$-0.07~dex in the more metal-poor KIC\,8145677 to $\sim$-0.12~dex in the more metal-rich stars tested. We did also check the MPIA tool\footnote{\url{https://nlte.mpia.de/gui-siuAC_secE.php}} in all stars for O, Si, Ca, Cr, Mn, and Co \citep[][]{Mashonkina:2007,Bergemann:2010b,Bergemann:2010,Bergemann:2013,Bergemann:2019,Bergemann:2021,Voronov:2022}. Again, NLTE departures are mild ($\lesssim$~0.05~dex), except for Co in KIC\,8145677, whose NLTE departure increases A(Co) by 0.19~dex, or $\sim$2-$\sigma$. For Al, the expected NLTE correction in the 669-nm doublet measured in this work is assumed to be of the order of -0.05~dex in our stellar parameter range \citep[][]{Nordlander:2017}.

\subsection{Grid-based modelling issues}
\label{sec:resultages}

As in Paper~I, we performed grid-based modelling runs including and not including Gaia parallaxes. Given the differences in ages found for a few stars and their sensitivity to the parallax zero-point, we chose to adopt the ages calculated without taking parallaxes into account as their nominal ages. Here we will highlight two stars whose analysis presented a few issues.

A star with a large difference in ages calculated with and without Gaia DR2 parallaxes is Group Four target KIC\,6940126. While the solution that does not take parallax into account gives an age of 8.4$^{+5.2}_{-3.3}$~Gyr, the adoption of its Gaia DR2 parallax pushes the age towards the 20~Gyr grid limit, giving 18.3$^{+1.2}_{-1.5}$~Gyr as the result. Neither of the two solutions is able to fit our estimated value of $\Delta \nu$ of 21.61~$\pm$~2.05~$\mu$Hz inside an 1-$\sigma$ interval. It is important to note that the oscillation power excess of KIC\,6940126 is near or above the Nyquist limit for \emph{Kepler}'s low-cadence observations, and, thus, the determination of $\nu_{\mathrm{max}}$ might be problematic and unreliable, as the peak frequency might not be sampled appropriately. Despite the $\sim$\,10~Gyr difference in age between both solutions, the choice of which solution to use has no impact in the resulting abundances calculated for this object, because the difference in their estimated log~g values is 0.003~dex.

\begin{figure}
	\includegraphics[width=\columnwidth]{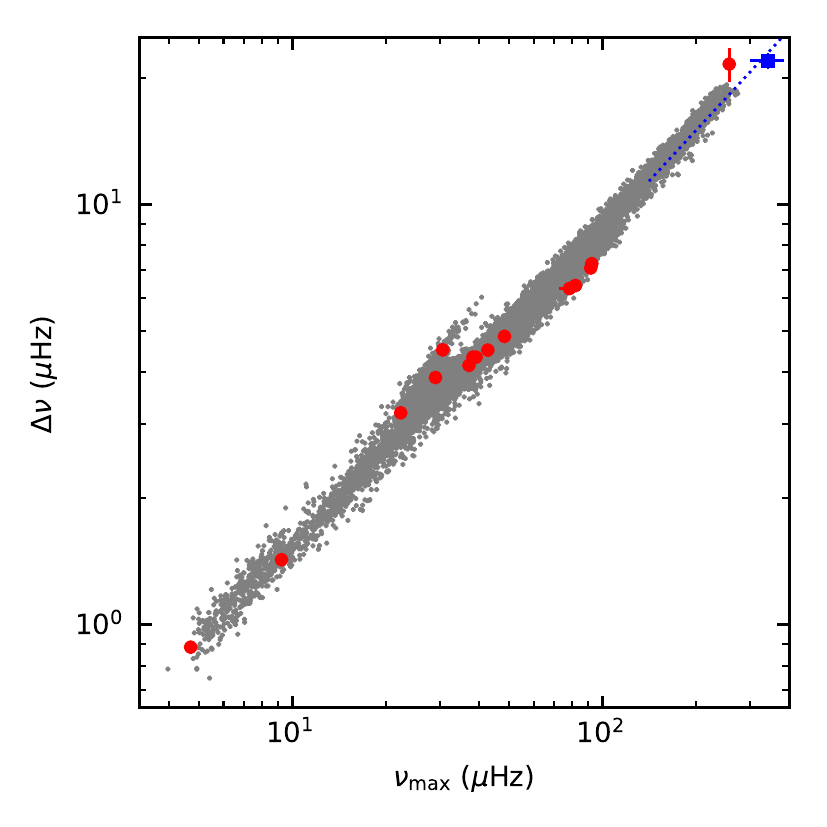}
    \caption{
    $\Delta \nu$ as function of $\nu_{\mathrm{max}}$.
    Red circles: values adopted for stars in this work (main-sequence star KIC\,6605673 not shown).
    Grey points: data from \citet[][]{Yu:2018}.
    The blue dotted line represents the linear fit for $\nu_{\mathrm{max}} >$~300~$\mu$Hz from \citet[][SYD pipeline]{Huber:2011b}.
    The blue square shows the \textsc{basta} solution for KIC\,6940126 when no asteroseismic parameters are used in the input data (see text for discussion).
    }
    \label{fig:numax_vs_dnu_kic694}
\end{figure}

The distance calculated for KIC\,6940126 in the solution using parallax is 1021$^{+8}_{-7}$~pc, in agreement with \citet[][]{BailerJones:2018} (1018$^{+18}_{-16}$~pc), and also with \citet[][]{BailerJones:2021}, which uses DR3 data (1012$^{+13}_{-12}$~pc). On the other hand, the solution that ignores parallax provides a distance of 1137$^{+62}_{-61}$~pc, resulting in a larger distance modulus, and, hence, an object with a larger intrinsic luminosity. The corresponding difference in the absolute \emph{K$_{s}$} magnitude estimated in each of the solutions is 0.24~mag. Interestingly, if a \textsc{basta} run is done \emph{without the asteroseismic parameters in the input set}, a solution taking the DR2 parallax value from Table~\ref{tab:tab1} yields an age of 7.7$^{+4.4}_{-2.6}$~Gyr, a distance of 1000$^{+19}_{-18}$~pc, and $\Delta \nu$ = 22~$\pm$~1~$\mu$Hz, in agreement with the $\Delta \nu$ shown in Table~\ref{tab:tab1}. However, that solution results in $\nu_{\mathrm{max}}$ = 342~$\pm$~42~$\mu$Hz, and a corresponding log~g of 3.44~$\pm$~0.05~dex, 0.12~dex larger than the value adopted in Table~\ref{tab:atmparam}. That 0.12~dex difference in log~g might result in a non-negligible shift in the chemical abundances with respect to those abundances appearing in Table~\ref{tab:abundances1}. The fitted [M/H] is not sensitive to these changes in \textsc{basta} input, at least for this star. In short, the stellar age derived using parallax instead of $\Delta \nu$ and $\nu_{\mathrm{max}}$ is more compatible with its super-solar metallicity, and higher-cadence observations would be necessary for an asteroseismic confirmation. Also, when compared to asteroseismic data in the literature (see Fig.~\ref{fig:numax_vs_dnu_kic694}), the super-Nyquist $\nu_{\mathrm{max}}$ resulting from the non-seismic constrained \textsc{basta} run for KIC\,6940126 seems (by inspection) to have a better agreement with the overall $\Delta\nu$-$\nu_{\mathrm{max}}$ relation for red giants.

The grid-based modelling of KIC\,9002884, one of the $\alpha$-rich stars from \citet[][]{Martig:2015aa}, has given problematic results -- not only the age uncertainties are large with the 16th- and 84th-percentiles corresponding to 6.8 and 16.9~Gyr, but the age distribution is bimodal with a secondary peak at $\approx$8~Gyr. The ages adopted in this work are the medians of the \textsc{basta} posteriors, for KIC\,9002884 the median is 13.1~Gyr. If we use the Gaia DR2 parallax from Table~\ref{tab:tab1}, the median is similar, the uncertainties decrease, the secondary peak vanishes, and the estimated age becomes 13.4$^{+2.1}_{-1.5}$~Gyr. However, when its Gaia DR3 parallax is included in the input its age drops to 2.6$^{+0.5}_{-0.5}$~Gyr, not only a huge quantitative swing, but resulting in an entirely different interpretation about its nature from the qualitative point of view. We have undertaken three tests in {\sc basta} excluding $\Delta \nu$ and $\nu_{\mathrm{max}}$ information and using Gaia DR2 or DR3 parallaxes. Both yielded large uncertainties as expected, while the median ages for the DR2 and DR3 experiments are similar to those estimated when asteroseismic information is included alongside with the respective parallax values.

From the test described above it seems that grid-based modelling is sensitive to the parallax in this object. Looking at the fitted asteroseismic parameters may give us a clue: two {\sc basta} runs, one excluding parallax information, and another adopting DR2 parallax, give similar ages around 13~Gyr for KIC\,9002884 and result in a $\Delta \nu$ value of 0.90~$\mu$Hz near the observed value of 0.885~$\pm$~0.029~$\mu$Hz. Meanwhile, the {\sc basta} run that adopts the DR3 parallax estimates 0.82~$\mu$Hz for $\Delta \nu$. All runs result in the same value of $\nu_{\mathrm{max}}$ -- 4.70~$\mu$Hz. For constant $\nu_{\mathrm{max}}$ and T$_{\mathrm{eff}}$, stellar mass scales with the inverse of $\Delta \nu$, which is observed in the younger age (i.e., larger mass) estimated with the DR3 parallax.

It is important to note that the large swing in age seen for KIC\,9002884, from $\approx$13~Gyr to 2.6~Gyr, is likely due to the large difference between Gaia DR2 and DR3 parallaxes. The DR2 value of 0.3859~$\pm$~0.0203~mas listed in Table~\ref{tab:tab1} is 0.1121~mas larger than the DR3 parallax $\varpi =$~0.2738~$\pm$~0.0128~mas, corrected according to the prescription from \citet[][]{Lindegren:2021}. That difference is larger than 5-$\sigma$ if we consider the DR2 uncertainty, or larger than 8-$\sigma$ if the DR3 uncertainty is to be considered. Besides the possible systematic underestimation of the DR3 uncertainties suggested in Paper~I, inaccuracies in zero-point corrections or in the astrometric solutions themselves might be sources of error.

\begin{figure*}
	\includegraphics{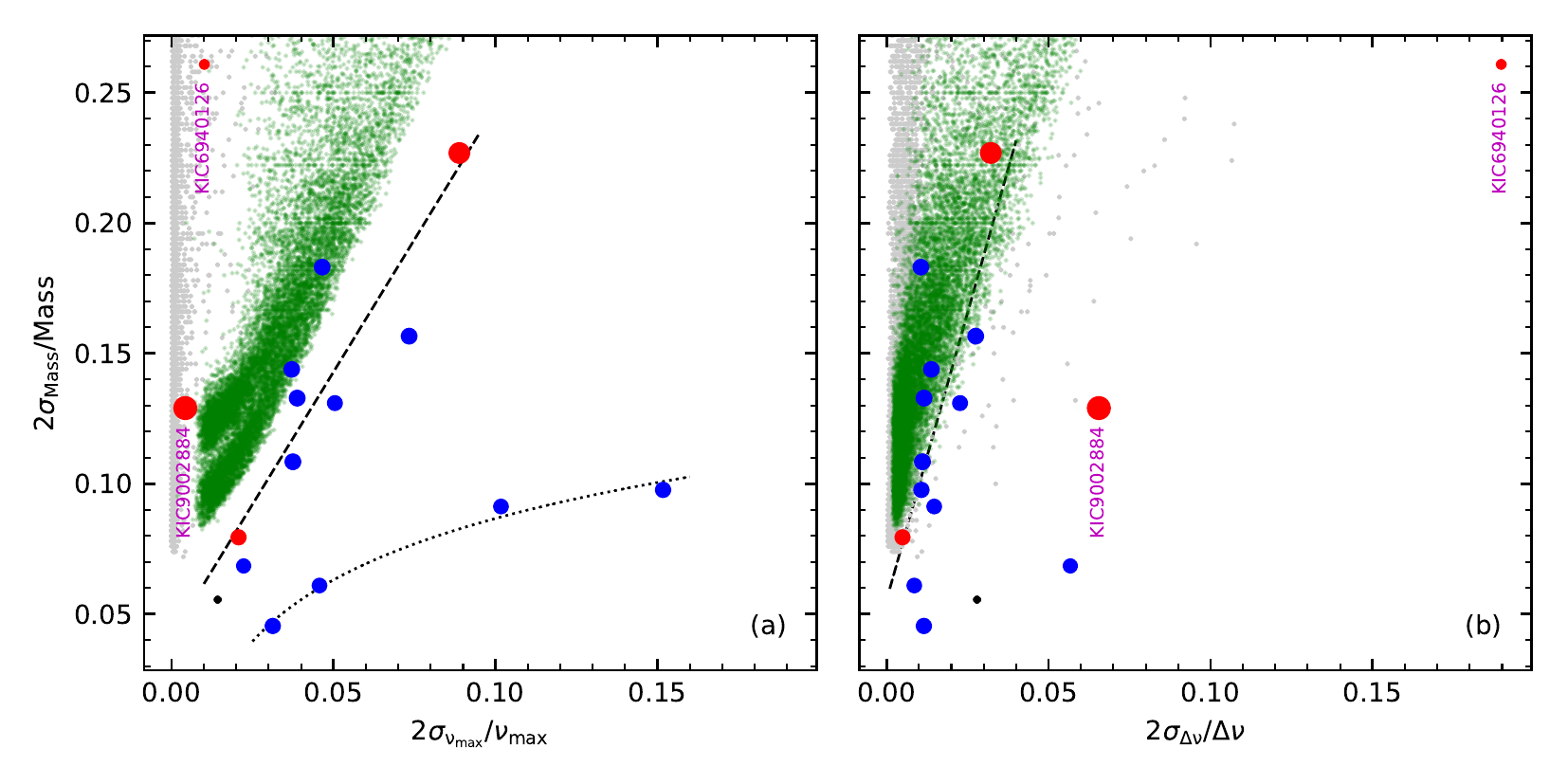}
    \caption{2-$\sigma$ (84th- minus 16th-percentile) fractional uncertainties in mass against 2-$\sigma$ fractional uncertainties in $\nu_{\mathrm{max}}$ (a) and 2-$\sigma$ fractional uncertainties in $\Delta \nu$ (b).
    Marker sizes scale linearly with radii.
    Colours indicate evolutionary stage.
    \emph{Black}: Main-Sequence.
    \emph{Red}: Red Giant Branch.
    \emph{Blue}: Core Helium Burning phase.
    Linear (dashed line) and logarithmic fits (dotted line) to our data are placed to guide the eye.
    Green dots are data from \citet[][]{Yu:2018}, while grey dots represent data from APOKASC-2 \citep[][]{Pinsonneault:2018}.
    Anomalous stars in this study are labelled (see text for discussion).}
    \label{fig:mass_unc_vs_numax_dnu_unc}
\end{figure*}

Given the importance of ages in reconstructing Galactic evolution, a fundamental question we raise is which solution should we trust? From our tests, the large swing in Gaia parallaxes from DR2 to DR3 is biasing the estimated ages. Hence, the (very old) age calculated without adopting parallaxes seems to be more accurate. However, the previously mentioned double peak in the age distribution and its large uncertainties (when no parallax is considered) still pose a problem for the analysis of this star, suggesting a lack of accuracy in at least one of the observables. As an additional check, we did compare the angular diameter derived with radii and distances from \textsc{basta} solutions with that calculated with the IRFM: 0.094~$\pm$~0.004~mas (the uncertainty is based on a conservative estimate of 4\%, see Table~\ref{tab:ap_irfm} for the other stars). The fitted surface gravity values differ by less than 0.005~dex, and the more massive DR3-based solution gives a larger stellar radius for KIC\,9002884. All three asteroseismic angular diameters calculated from the output of each run are inside the IRFM uncertainty, even if we consider 2\% instead of 4\% for the uncertainty estimation. Comparing the Gaia-based distances from \citet[][]{BailerJones:2018} and \citet[][]{BailerJones:2021}, which use data from DR2 and DR3, respectively, the distances derived by \citet[][]{BailerJones:2021} are $\approx$200~pc larger than the value of 2816$^{+173}_{-155}$~pc from \citet[][]{BailerJones:2018}. We have also found a larger distance in our DR3-based solution, 3349$^{+88}_{-135}$~pc, while the distances derived with the DR2 parallax and without parallax are basically the same: 2628$^{+32}_{-42}$~pc and 2636$^{+48}_{-47}$~pc, respectively. Given the discrepancy between Gaia DR2 and DR3 for this object, adopting ages calculated without parallax input for the whole sample seems to be the safest decision in this work, as it was done in Paper~I, despite resulting in larger error bars.

In Fig.~\ref{fig:mass_unc_vs_numax_dnu_unc} the relative uncertainties of mass as function of the relative uncertainties in the average asteroseismic observables are shown. The two targets discussed in the previous paragraphs are highlighted in the plots, and it can be seen that both deviate from the overall trend. In the case of relative $\nu_{\mathrm{max}}$ uncertainty, we can also see that the group of secondary RC stars, discussed in Section~\ref{sec:other}, follow a particular trend. Fig~\ref{fig:mass_unc_vs_numax_dnu_unc} strongly suggests that our method is able to yield masses with better precision when compared to the large-scale surveys shown in comparison, as it tends to result in more precise masses for the same level of precision in $\Delta \nu$ and $\nu_{\mathrm{max}}$.

\subsection{APOKASC young alpha-rich stars}
\label{sec:youngalpharich}

\citet{Martig:2015aa} investigated age-abundance relations using the first data release of APOKASC \citep{Pinsonneault:2014aa}. They identified 14 $\alpha$-rich stars younger than 6~Gyr (out of a sample of 241), seven of which we observed as part of this project, and which we discuss in more detail below. Their seismic-derived parameters as well as their $\alpha$ abundances calculated in this study are shown in Table~\ref{tab:martig}. Four of these stars are included in \citet[][]{Matsuno:2018aa}, whose [Fe/H] values are systematically lower by 0.2-0.3~dex than in this work.

\begin{table}
	\centering
	\caption{Fitted parameters and $\alpha$ abundances for young stars characterized as $\alpha$-rich by \citet{Martig:2015aa}. Here, the objects in the core He-burning phase (marked with an asterisk) have radii between 11 and 12~R$_{\odot}$. The remaining stars are ascending the RGB. The objects with high probability of belonging to binary systems \citep[][]{Jofre:2016aa} are marked with a dagger symbol.}
	\label{tab:martig}
	\begin{tabular}{rrrrr}
		\hline
		KIC & Age   & Mass          & Radius        & $[\alpha$/FeI$]$ \\
            & (Gyr) & (M$_{\odot}$) & (R$_{\odot}$) &              dex \\
        \hline
            \dag3455760  &  2.8$^{+0.5}_{-0.3}$ & 1.51$^{+0.05}_{-0.07}$ & 10.3$^{+0.1}_{-0.2}$ & 0.10 $\pm$ 0.09 \\
            *3833399  &  3.0$^{+0.6}_{-0.4}$ & 1.43$^{+0.10}_{-0.09}$ & 11.5$^{+0.3}_{-0.3}$ & 0.05 $\pm$ 0.06 \\
            \dag*5512910  &  2.8$^{+0.6}_{-0.5}$ & 1.39$^{+0.09}_{-0.11}$ & 11.1$^{+0.3}_{-0.3}$ & 0.11 $\pm$ 0.08 \\
            \dag9002884  & 13.1$^{+3.8}_{-6.3}$ & 0.92$^{+0.16}_{-0.06}$ & 26.9$^{+1.0}_{-0.7}$ & 0.24 $\pm$ 0.11 \\
            9761625  &  5.9$^{+3.5}_{-1.8}$ & 1.19$^{+0.13}_{-0.14}$ & 21.4$^{+0.9}_{-1.0}$ & 0.16 $\pm$ 0.07 \\
            \dag*10525475 &  2.9$^{+1.1}_{-0.5}$ & 1.42$^{+0.11}_{-0.15}$ & 11.1$^{+0.3}_{-0.5}$ & 0.13 $\pm$ 0.07 \\
            \dag*11823838 &  1.8$^{+0.3}_{-0.2}$ & 1.66$^{+0.07}_{-0.11}$ & 11.4$^{+0.2}_{-0.3}$ & 0.16 $\pm$ 0.08 \\
        \hline
	\end{tabular}
\end{table}

Of the seven stars in common with \citet[][]{Martig:2015aa}, five are indeed young (or seem to be young). One of them, KIC\,3833399, has solar [$\alpha$/Fe], and KIC\,5512910 sits on the boundary between $\alpha$-rich and $\alpha$-normal defined by the white line in Fig.~\ref{fig:afe_vs_feh}. Also, while most of the stars discussed in this Section seem \emph{visually} $\alpha$-rich in Fig.~\ref{fig:afe_vs_feh}, it is important to point out that only two have their [$\alpha$/Fe] differing from solar by more than 2-$\sigma$. Still, while adding this caveat that their status as $\alpha$-rich is disputed, we adopt the nomenclature for consistency with previous works in the literature. Finally, there are two stars which instead do not seem necessarily too young and we discuss them to follow.

In this work, KIC\,9761625 sits on the limit for 'young' stars as defined by \citet{Martig:2015aa} for both age and mass (6~Gyr and 1.2~M$_{\odot}$, respectively). In order to check whether the differences in the adopted input observables were relevant, we replaced our values of $\nu_{\mathrm{max}}$, $\Delta \nu$, T$_{\mathrm{eff}}$, and [M/H] with those published by \citeauthor{Martig:2015aa} for this object, as well as their adopted solar reference values. The grid-based modelling returned roughly the same age and mass as those shown in Table~\ref{tab:martig} ($\sim$6~Gyr and $\sim$1.2~M$_{\odot}$). If no model-based correction is applied to $\Delta \nu$ -- using the paramter \emph{dnuscal} instead of \emph{dnuSer} in \textsc{basta} -- the resulting mass is increased to 1.29~$\pm$~0.12~M$_{\odot}$. This mass value is within the error bar from \citeauthor{Martig:2015aa} and, interestingly, it is the same mass published by APOKASC-2, where a scaling relation correction was applied. The average seismic parameters adopted here for KIC\,9761625 have been taken from \citet{Yu:2018}, whose published (corrected) mass is 1.35~$\pm$~0.21 M$_{\odot}$. However, the T$_{\mathrm{eff}}$ adopted by \citeauthor{Yu:2018} is 200~K hotter than that shown in Table~\ref{tab:atmparam}, and a mass larger by $\sim$~0.1 M$_{\odot}$ in the scaling relation\footnote{Here we are using the scaling relation for a first order estimate.} could be attributed to a difference in temperature of that magnitude.

Despite being previously classified as 'young $\alpha$-rich' by \citet{Martig:2015aa} and '$\alpha$-normal' by \citet[][see discussion later in this subsection]{Hekker:2019}, KIC\,9002884 has an age estimation of 13.1$^{+3.8}_{-6.3}$~Gyr in this work, with $\alpha$-enhancement, i.e., its chemistry suggests that this object is indeed old ([$\alpha$/Fe] = +0.24 $\pm$ 0.11, see Table~\ref{tab:martig}). This star has a problematic age determination, already discussed in Section~\ref{sec:resultages}. From the chemical perspective, KIC\,9002884 displays enhancement in species such as Mg and Eu, seen in Fig.~\ref{fig:abundances} as the most enhanced green circle for both species. This could be an indication that this object would be an old member of the Thick Disc or the Bulge that currently inhabits the solar circle. However, asteroseismic data of higher quality than Kepler's is needed to constrain its fundamental parameters and to refine the analysis.

\subsubsection{Differences in \ion{Fe}{I} and \ion{Fe}{II} yield distinct interpretations}

In their near-infrared analysis, \citet{Hekker:2019} classified the stars in their sample as either 'young $\alpha$-rich', 'old $\alpha$-rich' or '$\alpha$-normal'. Here, [$\alpha$/Fe] is defined as the average of [X$_i$/Fe], X$_i$ = (Mg, Si, Ca, Ti), the same definition employed by \citeauthor{Hekker:2019}. In their work, KIC\,5512910, KIC\,10525475 and KIC\,11823838 were classified as 'young $\alpha$-rich'. The other four stars in Table~\ref{tab:martig} were classified by \citeauthor{Hekker:2019} as '$\alpha$-normal', i.e., young stars that do not display $\alpha$-enhancement.

Among those $\alpha$-normal in the \citet[][]{Hekker:2019} sample is KIC\,9002884, for which we obtain [$\alpha$/Fe]~$=$~0.24 in contrast with [$\alpha$/Fe]~$=$~0.07 measured in their work. It is interesting to point out, however, that in the case of KIC\,9002884 both studies give similar [$\alpha$/H] values, $-$0.08~dex in case of \citeauthor{Hekker:2019}, $-$0.12~dex in this work, with the [Fe/H] calculated here being 0.21~dex lower, which pushes [$\alpha$/Fe] to 0.24~dex.

\begin{figure}
	\includegraphics[width=\columnwidth]{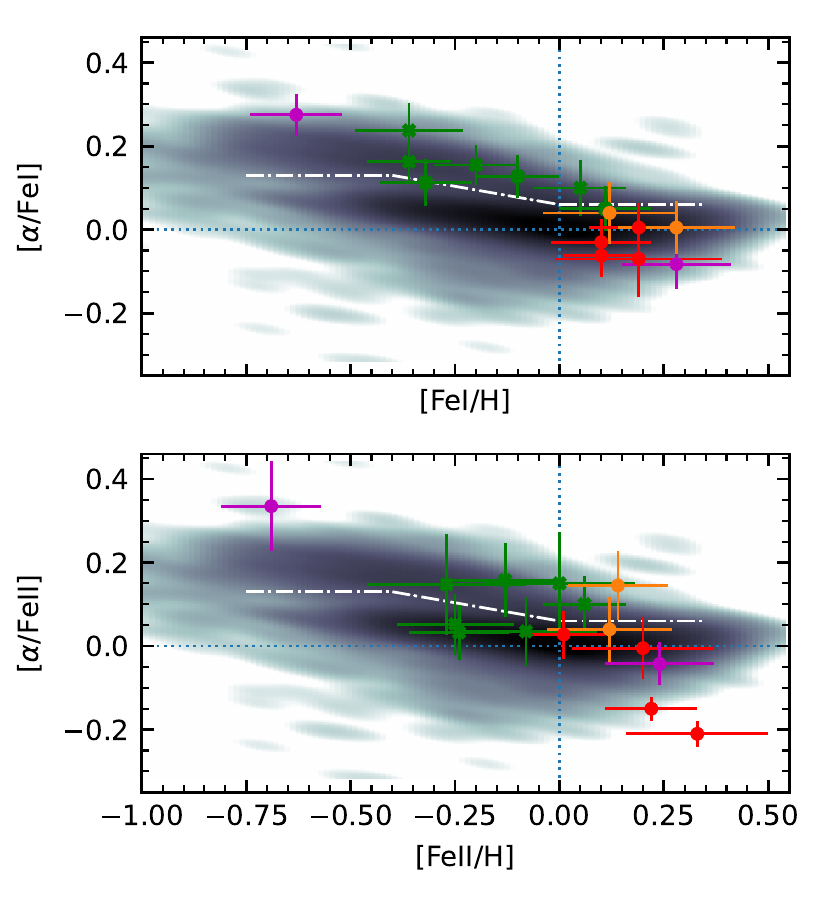}
    \caption{
    [$\alpha$/Fe] as function of [Fe/H] normalised for both \ion{Fe}{I} (top) and \ion{Fe}{II} (bottom).
    Point colours follow Fig.~\ref{fig:sample}, and density maps are described in Fig.~\ref{fig:zn_corr}.
    The white dash-dotted line represents the border between $\alpha$-rich and $\alpha$-normal adopted in this study. The abundance densities shown in greyscale are [Mg,Si,Ca,Ti/Fe] from APOGEE-2 \citep[][]{Jonsson:2020}.
    }
    \label{fig:afe_vs_feh}
\end{figure}

It must be noted that \citet[][]{Hekker:2019} Fe measurements are from \ion{Fe}{I}, while in this study we did measure both \ion{Fe}{I} and \ion{Fe}{II} lines. Because of our adoption of asteroseismic log~g, which is assumed to be both more accurate and more precise than the spectroscopic one, we did not enforce ionisation balance. That means that \ion{Fe}{I} and \ion{Fe}{II} abundances are not required to provide identical values, but, as shown in Fig.~\ref{fig:afe_vs_feh}, our results allow for different \emph{qualitative} interpretations for Fe (and $\alpha$) abundances depending on the ionisation state chosen for this chemical element.  If we choose \ion{Fe}{I}, we find that all stars are consistent with being $\alpha$-rich -- above the white dash-dotted line in Fig.~\ref{fig:afe_vs_feh}. Two of them, KIC\,3833399 and KIC\,5512910, have their data points lying near the boundary between $\alpha$-normal and $\alpha$-rich, which was based on the APOGEE-2 distribution also shown in Fig.~\ref{fig:afe_vs_feh}.

When \ion{Fe}{II} is taken as representative for Fe, however, the interpretation changes. Three stars -- KIC\,5512910, KIC\,9761625 and KIC\,11823838 -- are now on the $\alpha$-normal region, two of them in disagreement with the results from \citeauthor{Hekker:2019}. In some cases there is substantial disagreement between abundances from both Fe species, although the uncertainties overlap (see Table~\ref{tab:abundances1}).

All these stars except KIC\,3833399 were included in the study from \citet{Jofre:2016aa} that estimated the probabilities p$_b$ of these objects being in binary systems. The targets shown in Table~\ref{tab:martig} and studied by \citeauthor{Jofre:2016aa} have p$_b$~=~1, except KIC\,9761625 (p$_b$~=~0.06). This star is one of the three objects in this subset that look '$\alpha$-normal', i.e., below the white dash-dotted line in Fig.~\ref{fig:afe_vs_feh} if $\alpha$ abundances are normalised by \ion{Fe}{II} instead of \ion{Fe}{I}. The combination of $\alpha$-enhancement and binarity would give a strong hint on these stars being evolved blue stragglers, as the rate of binary systems is much higher in blue stragglers than in 'normal' objects, and a past mass accretion event would mask an old, $\alpha$-rich red giant as a young $\alpha$-rich star (\citet[][]{Yong:2016aa} and references therein, but see \citet[][]{Jofre:2023}).

\begin{figure}
	\includegraphics[width=\columnwidth]{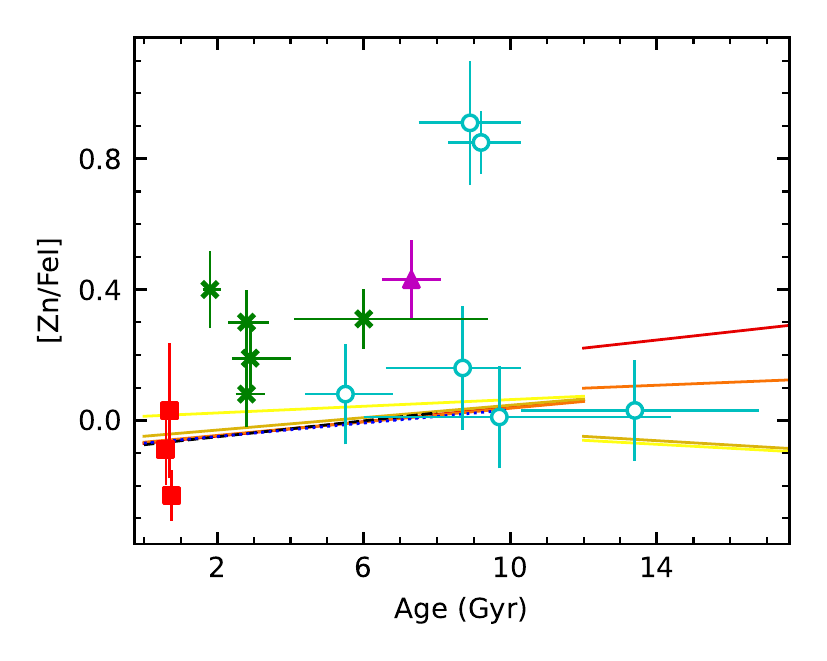}
    \caption{
    [Zn/Fe] as function of stellar age.
    Marker colours are the same as in Fig.~\ref{fig:sample}.
    The magenta triangle is KIC\,6605673, a dwarf.
    Cyan open circles are metal-poor objects from Paper~I.
    Solid lines show [Zn/Fe] vs. age linear fits for different metallicity bins calculated by \citet{Lin:2020} for main-sequence turnoff stars and subgiants using GALAH abundances \citep{Buder:2018aa}. Line colours represent the [Fe/H] bins adopted by \citeauthor{Lin:2020}, [-0.5, -0.1]: red, [-0.1, 0.0]: orange, [0.0, 0.1]: gold, and [0.1, 0.5]: yellow.
    The black dashed line represents the linear fit calculated by \citet{Spina:2016aa} using solar twins.
    }
    \label{fig:znfe}
\end{figure}

\subsubsection{Going beyond $\alpha$ abundances}

While the information given by the $\alpha$ abundances is inconclusive on the status of this subset, Zn could be useful to shed some light on the discussion. [Zn/Fe] is plotted as function of stellar age in Fig.~\ref{fig:znfe} for all stars with Zn measurements in this study and in Paper~I. The plot also shows linear fits made by \citet{Spina:2016aa} and \citet{Lin:2020} in their age-abundance studies for solar-type and subgiant stars, respectively. The stars from Table~\ref{tab:martig} (green crosses in Fig.~\ref{fig:znfe}) display a systematic excess of Zn that is also detected for Mg in the age-[Mg/Ni] relation from Fig.~\ref{fig:zn_corr}. This excess seems consistent with the interpretation that their chemistry is indicative of older ages than derived from their masses. In both cases, our young metal-rich RC stars (to be discussed in Section~\ref{sec:other}, and here assumed to be chemically normal stars), can be used to anchor our metallicity scale against the APOGEE-2 cloud from Fig.~\ref{fig:zn_corr} and the linear fits from Fig.~\ref{fig:znfe}. Given that both Mg and Zn are tracers from core-collpase supernovae \citep{Nomoto:2013aa,Kobayashi:2020}, from their chemistry alone it is unlikely that these stars are young thin disc objects. Also, binarity has been detected by radial velocity monitoring for most of these objects, as mentioned in the previous paragraph, hinting on the possibility of past mass accretion events for each of these red giants, making them look younger than they would actually be. The fact that their ages are possibly masked by past mass accretion events has been already pointed by \citet[][]{Jofre:2016aa}.

However, for KIC\,5512910 and KIC\,11823838 the interpretation becomes complicated if we consider the Mg abundance normalised by \ion{Fe}{I}. Their [Mg/Fe] abundance ratios fall in the intermediate region between normal- and enhanced-Mg in Fig.~\ref{fig:abundances} -- the same happens for Si, Ca, and Ti. Considering the error bars these two objects could be either normal or enhanced for these species. Also, when the orbits are taken into account (see Table~\ref{tab:kinematics}) the scenario becomes even more unclear. If we assume that the stars being discussed here in Section~\ref{sec:youngalpharich}, with the exception of KIC\,9002884, are old $\alpha$-rich objects that did experience mass accretion events in the past, one would expect thick disc-like orbits. Nevertheless, our orbital modelling from Gaia data shows that three of them -- KIC\,3455760, KIC\,3833399 and KIC\,11823838  -- have |Z$_{\mathrm{max}}$| < 0.5~kpc, and action J$_{\mathrm{z}} <$~5~kpc~km~s$^{-1}$, staying close to the Galactic plane. Of these three, KIC\,3833399 in particular does not seem to fit in the \emph{false-young} $\alpha$-rich categorisation with the data we currently have available, due to its low $\alpha$ abundance. It is one of the two stars near the $\alpha$-normal/rich boundary in Fig.~\ref{fig:afe_vs_feh} ([$\alpha$/FeI] = 0.05), and it has a thin disc orbit. As previously mentioned, KIC\,3833399 does not appear in \citet[][]{Jofre:2016aa} analysis of binarity. \citet[][]{Jofre:2023} has radial velocity measurements of KIC\,3833399, and the target was flagged as non-binary in their study. Among the stars in common that were classified as young $\alpha$-rich by \citet[][]{Hekker:2019}, from our results only KIC\,10525475 seems to meet the requirements to be in our definition of the 'false-young' category -- that is, $\alpha$-rich giant \emph{and} with stellar mass larger than 1.2~M$_{\odot}$ \emph{and} with a heated thick disc-like orbit \emph{and} being part of a binary system \citep[although binarity might not be decisive, see][]{Jofre:2023}. This definition assumes that the 'false-young' giants have formed before the thin disc, and underwent mass accretion from a companion.

On the other hand, when the N/C ratios of these objects published by \citet{Hekker:2019} are taken into account, none of these stars show unusual N enhancement expected from the first dredge-up in higher mass stars \citep[see also][]{Martig:2016aa} -- their published N/C values range between 0.4~$\pm$~0.1 and 0.8~$\pm$~0.3. This would suggest a 'false-young' scenario for these young $\alpha$-rich giants. Also, when the z-component of the angular momentum L$_{\mathrm{z}}$ is taken into account and compared to [Mg/Fe], as done by \citet[][fig.~8]{Ness:2019}, it is expected that stars in the L$_{\mathrm{z}}$ and age intervals of the subset discussed here -- $\sim$ 1500 kpc km s$^{-1}$ and $<$6~Gyr, respectively -- would have near-solar [Mg/Fe], which is not the case for any of the suspected young and $\alpha$-rich stars discussed in this work. That is, from their Mg abundances and angular momenta we should expect them to be older ($\gtrsim$10~Gyr), but the caveat here is that the study from \citeauthor{Ness:2019} was carried out with a selected sample of low-$\alpha$ stars.

\subsection{Old metal-rich stars}
\label{sec:omr}

In this study we define as 'old and metal-rich' those stars whose [Fe/H] is larger than zero \emph{and} whose age is larger than 10~Gyr. Two of the program stars fall into this category: KIC\,6936796 and KIC\,7595155. Both targets are Red Clump objects \citep[][also confirmed by our grid-based modelling analysis]{Ting:2018,Yu:2018}. In both objects the super-solar metallicity determined in this study confirms results previously published in the literature regarding their overall chemical composition \citep[][]{Casagrande:2014aa,Ness:2015aa,Hawkins:2016,Xiang:2019,Jonsson:2020}.

Our adopted age for KIC\,6936796 is 14$^{+2.8}_{-2.8}$~Gyr (see Table~\ref{tab:omr}). When its (corrected) Gaia DR2 parallax is taken into consideration, the estimated age jumps to 18.2$^{+1.3}_{-1.4}$~Gyr. The asteroseismic observables $\Delta \nu$ and $\nu_{\mathrm{max}}$ estimated by \textsc{basta} solutions have better agreement with the inputs listed in Table~\ref{tab:tab1} (the \emph{observed} inputs) when the parallax is \emph{not} taken into consideration. Regarding $\Delta \nu$, the value fitted when the DR2 parallax is employed as input diverges from the observed value by more than 1-$\sigma$. The adopted value for stellar mass, 0.83$^{+0.07}_{-0.06}$~M$_{\odot}$, is in agreement with those published both by APOKASC-2 and \citet{Yu:2018} when taking uncertainties into account. Interestingly, in APOKASC-2 the estimated age for KIC\,6936796 is lower, 10.8$^{+4.2}_{-3.0}$~Gyr, despite its values of T$_{\mathrm{eff}}$ (4562~K) and metallicity ([Fe/H] = +0.28) being similar to those shown in Table~\ref{tab:atmparam}.

The chemistry of KIC\,6936796 displays some [Mg,Si/Fe] enhancement and solar [O,Ca,Cr,Ni/Fe] (within uncertainties), but subsolar [Ti/Fe], as well as deficiency of [V,Mn/Fe], while [Co/Fe] is above solar. Among the neutron-capture elements, the deficiency of Sr, Y, Zr and Ba is also noticeable, while Ce and Nd are solar, and a small deficiency is measured for Eu. \citet[][]{Hawkins:2016} did measure several light and Fe-peak species for this star using asteroseismic log~g, spectrophotometric T$_{\mathrm{eff}}$ and infrared spectrum. Their [X/Fe] abundance ratios are solar (under an 1-$\sigma$ interval) for several species also measured in this work (O, Mg, Si, Ti, Mn, Co). Meanwhile, the values from \citeauthor{Hawkins:2016} for Ca, V and Cr are below solar, and their [Ni/Fe] is slightly above solar. In both studies KIC\,6936796 is Al-enhanced with respect to the Sun by 0.2-0.3~dex. The differences between both studies are likely due to sensitivities of the abundance ratios to different sets of atmospheric parameters. KIC\,6936796 has measurements for all the six 'classic' light-s (ls) and heavy-s (hs) species, (Sr, Y, Zr) and (Ba, La, Ce). Its high [hs/ls] ratio (0.36~dex) is indicative of enrichment by low mass ($\lesssim$2~M$_{\odot}$) AGB stars at solar metallicity \citep[][]{Karakas:2016aa}, which is compatible with KIC\,6936796 [Fe/H] uncertainty. That corresponds to a timescale of at least 1.2-2.9~Gyr \citep[][]{Karakas:2014aa}, which might suggest that the age of KIC\,6936796 would be lower than the age of the Universe, but still compatible with our definition of 'old'.

\begin{table}
	\centering
	\caption{Fitted stellar parameters for the old metal-rich stars. Last column displays their estimated masses at the start of the main-sequence, assuming the mass loss parameter $\eta$ = 0.3.}
	\label{tab:omr}
	\begin{tabular}{lrrrr}
		\hline
		KIC & Age & Mass & Radius & Initial Mass \\
            & (Gyr) & (M$_{\odot}$) & (R$_{\odot}$) & (M$_{\odot}$) \\
        \hline
            6936796 & 14.0$^{+2.8}_{-2.8}$ & 0.83$^{+0.07}_{-0.06}$ & 11.4$^{+0.4}_{-0.3}$ & 1.00$^{+0.06}_{-0.05}$ \\
            7595155 & 14.9$^{+2.3}_{-2.3}$ & 0.84$^{+0.06}_{-0.05}$ & 10.0$^{+0.3}_{-0.2}$ & 1.00$^{+0.05}_{-0.04}$ \\
        \hline
	\end{tabular}
\end{table}

KIC\,7595155 displays solar [X/Fe] for Si, Mn, Ba, La, Ce, and Nd. Apart from these species and the enhanced elements O, Mg, Al and Co, all other [X/Fe] are subsolar in this giant. It seems to be $\alpha$-normal in Fig.~\ref{fig:afe_vs_feh}, if [$\alpha$/Fe] normalisation is done against \ion{Fe}{I}. \citeauthor{Hawkins:2016} report a much higher [Fe/H] of 0.50~dex -- compared to 0.28~$\pm$~0.14 in this work, and most of their [X/Fe] are around the solar value. The adopted age for KIC\,7595155 is 14.9$^{+2.3}_{-2.3}$~Gyr. At its metallicity the [hs/ls] from AGB contribution is not expected to be super-solar \citep[][]{Karakas:2016aa}. We were not able to measure its Sr and Y abundances, and if we take Zr as proxy for ls, its [hs/Zr] is too high when compared to the model expectations (0.24~dex).

KIC\,6936796 and KIC\,7595155 have similar stellar masses (see Table~\ref{tab:omr}), and the 0.84~M$_{\odot}$ value estimated for KIC\,7595155 is also in agreement with masses published by \citet{Yu:2018} and APOKASC-2. Based on the adopted mass loss parameter $\eta$~=~0.3, we estimate that their initial masses are 1~M$_{\odot}$ with \textsc{basta}. Super metal-rich stars with one solar mass at their zero-age main-sequence are expected to live for at least the age of the Universe \citet[][]{Karakas:2014aa}.

Fig.~\ref{fig:sample} shows that this pair of objects is indeed at one edge of the distribution of CHeB stars (black points). In Fig.~\ref{fig:sample}(a) and Fig.~\ref{fig:sample}(c) the (current) stellar mass correlates positively with $\nu_{\mathrm{max}}^{0.75}$/$\Delta\nu$, and KIC\,6936796 and KIC\,7595155 can be seen towards the low-mass end of the RC population. That is, when we inspect the information directly from $\Delta \nu$ and $\nu_{\mathrm{max}}$, these stars are expected to be low mass.

One hypothesis to explain their old ages and metal rich chemistry is to consider the possibility of radial migration. From the assumption that the Galactic disc has a tight age-radius-metallicity relation \citep[see, e.g., fig.~6 of][]{Schonrich:2009aa}, it is expected that super metal-rich stars inhabiting the solar circle have migrated outwards -- it is believed that the Sun itself has formed closer to the Galactic centre \citep[][]{Wielen:1996}. Objects such as KIC\,6936796 and KIC\,7595155 could be escapees from the bulge or the inner disc. Our kinematic analysis shows that both stars are solar circle dwellers with thin disc orbits (see Table~\ref{tab:kinematics}), having |Z$_{\mathrm{max}}$| $\approx$ 0.4~kpc and low (<~0.2) eccentricity. A non-zero fraction of old stars displays in-plane, solar circle orbits \citep[see][]{Casagrande:2016aa,Miglio:2021}, as they would have had enough time to circularise after migration \citep['churning',][]{Sellwood:2002,Schonrich:2009aa}.

\begin{table*}
	\centering
	\caption{The results from our orbital modelling. From right to left, perigalactic and apogalactic radii in Galactocentric coordinates, orbital energy, eccentricity, the maximum displacement from the Galactic plane $|$Z$_{\mathrm{max}}|$, and the actions J$_{\mathrm{r}}$, J$_{\mathrm{z}}$, and J$_{\mathrm{\phi}}$, here labelled as the z-component of the angular momentum L$_{\mathrm{z}}$. Uncertainties marked as '0.00' represent values smaller than 0.005.
	}
	\label{tab:kinematics}
	\begin{tabular}{lrrrrcrrr}
		\hline
		Star & R$_{\mathrm{p}}$ & R$_{\mathrm{a}}$ &            Energy & Ecc & $|$Z$_{\mathrm{max}}|$ & L$_{\mathrm{z}}$ & J$_{\mathrm{r}}$ & J$_{\mathrm{z}}$ \\
     &              kpc &              kpc & km$^{2}$ s$^{-2}$ &     & pc & kpc km s$^{-1}$ & kpc km s$^{-1}$ & kpc km s$^{-1}$ \\
        \hline
 KIC2845610 & 6.48$^{+0.01}_{-0.01}$ &  8.27$^{+0.00}_{-0.00}$ & -1.62e+05 & 0.121$^{+0.001}_{-0.001}$ &  145$^{+  1}_{-  1}$ &  1693$^{+ 1}_{- 1}$ &  18.1$^{+ 0.2}_{- 0.2}$ &  0.7$^{+0.0}_{-0.0}$ \\
 KIC3455760 & 5.90$^{+0.01}_{-0.02}$ &  8.09$^{+0.00}_{-0.00}$ & -1.65e+05 & 0.157$^{+0.001}_{-0.001}$ &  176$^{+  2}_{-  2}$ &  1589$^{+ 2}_{- 2}$ &  28.8$^{+ 0.5}_{- 0.4}$ &  1.1$^{+0.0}_{-0.0}$ \\
 KIC3833399 & 5.06$^{+0.01}_{-0.01}$ &  8.13$^{+0.00}_{-0.00}$ & -1.68e+05 & 0.233$^{+0.001}_{-0.001}$ &  382$^{+  4}_{-  4}$ &  1446$^{+ 1}_{- 1}$ &  60.2$^{+ 0.3}_{- 0.3}$ &  4.5$^{+0.1}_{-0.1}$ \\
 KIC5512910 & 5.57$^{+0.05}_{-0.05}$ &  8.22$^{+0.07}_{-0.06}$ & -1.65e+05 & 0.192$^{+0.006}_{-0.006}$ & 1169$^{+ 70}_{- 61}$ &  1508$^{+10}_{- 9}$ &  40.5$^{+ 2.6}_{- 2.4}$ & 30.1$^{+2.7}_{-2.3}$ \\
 KIC5707338 & 6.14$^{+0.03}_{-0.03}$ &  9.28$^{+0.03}_{-0.03}$ & -1.60e+05 & 0.203$^{+0.004}_{-0.004}$ &  370$^{+  8}_{-  9}$ &  1726$^{+ 5}_{- 5}$ &  52.9$^{+ 1.9}_{- 1.8}$ &  3.6$^{+0.1}_{-0.1}$ \\
 KIC6605673 & 8.11$^{+0.00}_{-0.00}$ &  9.34$^{+0.02}_{-0.02}$ & -1.53e+05 & 0.070$^{+0.001}_{-0.001}$ &  770$^{+  5}_{-  5}$ &  2004$^{+ 2}_{- 2}$ &   6.7$^{+ 0.2}_{- 0.2}$ & 13.2$^{+0.1}_{-0.2}$ \\
 KIC6634419 & 6.99$^{+0.13}_{-0.19}$ &  8.66$^{+0.20}_{-0.13}$ & -1.59e+05 & 0.106$^{+0.024}_{-0.016}$ &  161$^{+ 39}_{- 24}$ &  1805$^{+ 6}_{- 9}$ &  14.7$^{+ 7.4}_{- 4.2}$ &  0.9$^{+0.4}_{-0.2}$ \\
 KIC6936796 & 7.50$^{+0.01}_{-0.01}$ &  8.97$^{+0.01}_{-0.01}$ & -1.56e+05 & 0.090$^{+0.001}_{-0.001}$ &  440$^{+  5}_{-  5}$ &  1897$^{+ 2}_{- 2}$ &  10.8$^{+ 0.2}_{- 0.2}$ &  5.2$^{+0.1}_{-0.1}$ \\
 KIC6940126 & 5.45$^{+0.04}_{-0.04}$ &  8.21$^{+0.00}_{-0.00}$ & -1.66e+05 & 0.202$^{+0.004}_{-0.003}$ &  277$^{+  4}_{-  5}$ &  1524$^{+ 6}_{- 7}$ &  46.5$^{+ 1.5}_{- 1.4}$ &  2.5$^{+0.1}_{-0.1}$ \\
 KIC7595155 & 5.60$^{+0.02}_{-0.01}$ &  7.99$^{+0.00}_{-0.00}$ & -1.67e+05 & 0.176$^{+0.001}_{-0.001}$ &  499$^{+  9}_{-  8}$ &  1523$^{+ 3}_{- 2}$ &  34.9$^{+ 0.4}_{- 0.5}$ &  7.3$^{+0.2}_{-0.2}$ \\
 KIC8145677 & 2.70$^{+0.02}_{-0.02}$ &  9.31$^{+0.03}_{-0.03}$ & -1.71e+05 & 0.550$^{+0.004}_{-0.003}$ & 1091$^{+ 16}_{- 14}$ &   971$^{+ 4}_{- 5}$ & 303.8$^{+ 4.3}_{- 3.3}$ & 22.6$^{+0.5}_{-0.4}$ \\
 KIC9002884 & 6.95$^{+0.05}_{-0.05}$ &  8.42$^{+0.07}_{-0.06}$ & -1.60e+05 & 0.096$^{+0.002}_{-0.002}$ & 1402$^{+ 80}_{- 69}$ &  1718$^{+10}_{-10}$ &  10.9$^{+ 0.5}_{- 0.5}$ & 41.0$^{+3.6}_{-3.0}$ \\
 KIC9266192 & 6.49$^{+0.01}_{-0.01}$ &  8.66$^{+0.01}_{-0.01}$ & -1.61e+05 & 0.143$^{+0.001}_{-0.001}$ &  292$^{+  2}_{-  2}$ &  1728$^{+ 1}_{- 1}$ &  25.7$^{+ 0.3}_{- 0.3}$ &  2.6$^{+0.0}_{-0.0}$ \\
 KIC9761625 & 5.28$^{+0.08}_{-0.05}$ &  8.25$^{+0.03}_{-0.03}$ & -1.66e+05 & 0.220$^{+0.003}_{-0.005}$ & 1787$^{+130}_{-109}$ &  1427$^{+ 7}_{- 5}$ &  51.1$^{+ 1.4}_{- 1.7}$ & 58.2$^{+5.8}_{-4.9}$ \\
KIC10525475 & 6.23$^{+0.02}_{-0.02}$ &  8.68$^{+0.02}_{-0.01}$ & -1.61e+05 & 0.164$^{+0.001}_{-0.001}$ &  791$^{+ 12}_{- 11}$ &  1673$^{+ 4}_{- 3}$ &  32.9$^{+ 0.5}_{- 0.5}$ & 14.8$^{+0.4}_{-0.3}$ \\
KIC11823838 & 5.13$^{+0.06}_{-0.07}$ & 13.01$^{+0.21}_{-0.18}$ & -1.50e+05 & 0.434$^{+0.011}_{-0.010}$ &  463$^{+ 21}_{- 16}$ &  1756$^{+13}_{-11}$ & 283.5$^{+17.0}_{-14.2}$ &  3.2$^{+0.1}_{-0.1}$ \\
        \hline
	\end{tabular}
\end{table*}

Another possibility (which is not mutually exclusive with radial migration) is regarding the mass loss rate. These two objects might have suffered from some unusual extreme mass loss event, making our adopted $\eta$ = 0.3 for the Reimers formula \citep[][]{Reimers:1975} an underestimation. This value is based on the open cluster results from \citet[][]{Miglio:2012} and it is used as standard in the BaSTI grid \citep[][]{Hidalgo:2018}. A larger value of $\eta$ implies that the initial masses of KIC\,6936796 and KIC\,7595155 would be larger than 1~M$_{\odot}$, hence their ages would be lower. Also, it is possible that the adopted mass loss parameter might be underestimated for metal-rich stars in the lower mass end.

\citet[][]{Handberg:2017} found a RC giant, KIC\,4937011, that has potentially suffered from enhanced mass loss in the open cluster NGC\,6819. However, as \citeauthor{Handberg:2017} point out, it might not be a cluster member, see also \citet[][]{Stello:2011}. \citet[][]{Handberg:2017} proposed a relation between its lower mass and its high Li abundance. Nevertheless, neither KIC\,6936796 or KIC\,7595155 were identified as Li-rich in our study.

\begin{table}
	\centering
	\caption{Runs made with PARAM for KIC\,7595155, adopting the same observables from Tables \ref{tab:tab1} and \ref{tab:atmparam}. The first column shows which model grid was adopted: R17 \citep[][]{Rodrigues:2017} or PARSEC \citep[][]{Bressan:2012aa}. In the second column the adopted value for the mass loss parameter $\eta$ is displayed, while the third column informs if the DR2 parallax value from Table~\ref{tab:tab1} was included. The last two columns show the respective results for stellar age, current mass, and distance. The run with parallax information did not converge.}
	\label{tab:param}
	\begin{tabular}{lrrrrr}
		\hline
		Grid & $\eta$ & Plx? & Age & Mass & Distance \\
		     &        &      & Gyr & M$_{\odot}$ & pc \\
        \hline
            R17    & 0.3 & No  & 11.3$^{+1.9}_{-1.4}$ & 0.98$^{+0.04}_{-0.04}$ & 1517$^{+39}_{-37}$ \\
            R17    & 0.4 & No  & 10.7$^{+1.8}_{-1.5}$ & 0.98$^{+0.04}_{-0.05}$ & 1517$^{+39}_{-37}$ \\
            R17    & 0.7 & No  &  7.8$^{+0.5}_{-0.6}$ & 0.98$^{+0.05}_{-0.04}$ & 1522$^{+38}_{-37}$ \\
            PARSEC & 0.3 & No  & 12.2$^{+2.4}_{-2.0}$ & 0.86$^{+0.07}_{-0.06}$ & 1446$^{+61}_{-61}$ \\
            R17    & 0.3 & Yes &                  ... &                    ... &                ... \\
        \hline
	\end{tabular}
\end{table}

We tested different mass loss rates and model grids using \textsc{param} \citep[][]{daSilva:2006,Rodrigues:2014}\footnote{Using the web tool available at \url{http://stev.oapd.inaf.it/cgi-bin/param}.}, and those runs are summarised in Table~\ref{tab:param}. A tentative run including the DR2 parallax from Table~\ref{tab:tab1} and $\eta$=0.3 was performed, but convergence was not achieved. The results with R17 indicate consistently lower ages in comparison with our adopted \textsc{basta}+BaSTI values, and the resulting \emph{current} mass being 0.14 M$_{\odot}$ ($\sim$~3-$\sigma$) higher. Interestingly, when the $\eta$ value extrapolated from \citet[][]{Tailo:2020} is adopted (0.7), the age of KIC\,7595155 drops to a value compatible with the thin disc (7.8$^{+0.5}_{-0.6}$~Gyr). Meanwhile, the run using the PARSEC grid gives a mass similar to that estimated with \textsc{basta}+BaSTI. The resulting age is younger than that adopted in Table~\ref{tab:omr}, although older than that derived with R17. The differences in ages may be attributed to differences in the helium prescription of each model. \textsc{param} also outputs an estimation of stellar distance. The run with PARSEC yields a distance closer to the geometric estimation from \citet[][]{BailerJones:2021} (1418$^{+24}_{-16}$~pc) in comparison with the runs adopting R17, as well as with respect to the distance published by \citet[][]{BailerJones:2018} (1459$^{+50}_{-47}$~pc). The distance calculated with \textsc{basta}+BaSTI (without parallax input) is 1425$^{+25}_{-22}$~pc.

When checking the 6676 entries of the APOKASC-2 sample, there are 2437 RC stars without the \emph{SeisUnc} flag, of which 23 are older than 10 Gyr. The median [Fe/H] of these 23 objects is +0.11 dex. Applying the cut-off by mass, from the 74 RC stars with masses $<$ 0.85~M$_{\odot}$, there is an anticorrelation (Pearson $\rho$ = -0.53, p-value = 5 $\times$ 10$^{-7}$) between [Fe/H] and mass. Meanwhile, in APOKASC-2 there are no super metal-rich RGB stars less massive than 0.85~M$_{\odot}$. In the RGB, the lower mass limit in the APOKASC-2 sample is dominated by upper-RGB objects ($\nu_{\mathrm{max}}$~$<$~10~$\mu$Hz), and scales mildly with [Fe/H], while the number of RC stars less massive than the lower RGB limit grows with metallicity.

This trend might suggest a metallicity dependent mass loss, which is also suggested by globular clusters in a more metal-poor regime. However, as stated in \citet[][]{Pinsonneault:2018}, the APOKASC-2 results for RC stars must be treated with caution, due to uncertainties in their method related to this evolutionary stage and lack of calibration data in the relevant region of the parameter space. In addition to that, the asteroseismic results from \citet[][]{Miglio:2012}, \citet[][]{Handberg:2017} and \citet[][]{Miglio:2021} indicate a mild mass loss rate in the RGB. Also, there is evidence that low-mass RC stars may be the mass-stripped component from a binary system \citep[][]{Li:2022}.

Summing up, the hypothesis of radial migration should be kept open for these two particular stars. Their chemical profiles can be seen as ambiguous: low-$\alpha$, however enhanced Mg, but at the same time a high [hs/ls] ratio suggesting a timescale longer than that of typical $\alpha$-enhancement, at least for KIC\,6936796. Also, both their ages and masses are very sensitive to the adopted mass loss rate and model grid.

\subsection{Other targets}
\label{sec:other}

The remaining subset is composed by (i) the four secondary RC stars with strong non-radial mode damping (red markers in Fig.~\ref{fig:abundances}), (ii) another old RC object whose [Fe/H] is similar to that of the globular cluster 47~Tuc -- KIC\,8145677, and (iii) KIC\,6605673, a dwarf (see Table~\ref{tab:otherstars} for their fundamental parameters).

\begin{table}
	\centering
	\caption{Fundamental parameters and evolutionary phase of stars discussed in Section~\ref{sec:other}.}
	\label{tab:otherstars}
	\begin{tabular}{lrrrr}
		\hline
		KIC & Age & Mass & Radius & Phase \\
            & (Gyr) & (M$_{\odot}$) & (R$_{\odot}$) &  \\
        \hline
            2845610 &  0.8$^{+0.2}_{-0.1}$ & 2.41$^{+0.10}_{-0.12}$ &    9.7$^{+0.1}_{-0.2}$ &  RC \\
            5707338 &  0.6$^{+0.1}_{-0.1}$ & 2.64$^{+0.04}_{-0.08}$ &   10.6$^{+0.1}_{-0.1}$ &  RC \\
            6605673 &  7.3$^{+0.8}_{-0.8}$ & 1.08$^{+0.03}_{-0.03}$ & 1.63$^{+0.02}_{-0.02}$ &  MS \\
            6634419 &  0.7$^{+0.2}_{-0.1}$ & 2.46$^{+0.05}_{-0.10}$ &    9.6$^{+0.1}_{-0.1}$ &  RC \\
            6940126 &  8.4$^{+5.2}_{-3.3}$ & 1.15$^{+0.16}_{-0.14}$ &    3.9$^{+0.3}_{-0.2}$ & RGB \\
            8145677 & 13.5$^{+0.9}_{-1.4}$ & 0.73$^{+0.03}_{-0.02}$ &    8.9$^{+0.2}_{-0.2}$ &  RC \\
            9266192 &  0.7$^{+0.2}_{-0.2}$ & 2.56$^{+0.10}_{-0.15}$ &   10.6$^{+0.2}_{-0.2}$ &  RC \\
        \hline
	\end{tabular}
\end{table}

The group of secondary RC stars were included in the program because their observed power spectra look unusual: they all show significant mode broadening in the frequency power spectra, which is evidence for strong mode damping. Their masses estimated in this study are lower than those from \citet{Yu:2018} by 0.3-0.6~M$_{\odot}$. The masses derived here are close to the values published by \citet{Queiroz:2020}, without asteroseismic input, using the \textsc{StarHorse} code \citep{Queiroz:2018}. \textsc{StarHorse} agrees with our results for KIC\,5707338 and KIC\,9266192 within the error bars for masses. However, the log~g values estimated with \textsc{StarHorse} are systematically lower by $\sim$~0.1 dex for this subset of stars.

Despite the agreement with \textsc{StarHorse}, the solutions found with \textsc{basta} for the secondary RC stars are suboptimal. Differences larger than 1-$\sigma$ between input and output parameters are found in these objects. This includes a difference larger than 1-$\sigma$ between input and fitted $\nu_{\mathrm{max}}$ for KIC\,2845610, despite a (large) fractional uncertainty of 5\% in this observable. Also, for KIC\,6634419 there are no models with non-zero likelihood in a 2-$\sigma$ interval of the adopted $\nu_{\mathrm{max}}$ in the adopted BaSTI grid\footnote{That is, for the adopted set of (T$_{\mathrm{eff}}$, [M/H], $\Delta\nu$, $\nu_{\mathrm{max}}$, evolutionary phase, 2MASS $K_s$), no models are found if we constrain $\nu_{\mathrm{max}}$ in a $\pm$~2-$\sigma$ interval.}. In order to check for potential inaccuracies in the adopted asteroseismic parameters, we employed the code \textsc{PBjam}\footnote{\url{https://github.com/grd349/PBjam}} \citep[][]{Nielsen:2021} to re-extract $\Delta\nu$ and $\nu_{\mathrm{max}}$ from the Kepler light curves. The values extracted with \textsc{PBjam} differ from those shown in Table~\ref{tab:tab1} by more than 1-$\sigma$ in $\nu_{\mathrm{max}}$ -- for KIC\,5707338 and KIC\,9266192, and $\Delta\nu$ -- for KIC\,2845610 and KIC\,6634419. However, the set of asteroseismic parameters extracted with \textsc{PBjam} had no impact in the results -- all four stars are still young (0.5~Gyr $<$ Age $<$ 1.0~Gyr) and in the same mass range as previously calculated. Also, the difference in log~g is always lower than 0.03~dex.

From the chemical point of view this subset of young metal-rich thin disc RC stars appears normal, with two exceptions. They are relatively enhanced in Ba, particularly when compared to La and Ce, and one of them, KIC\,6634419, has [Ba/Fe] = 0.75 $\pm$ 0.22. This object also displays anomalous [Mg/Fe], [Si/Fe], and [Sc/Fe], and larger than average uncertainties in abundance ratios. There is evidence linking enhanced Ba abundances and young stellar ages \citep[see, e.g.,][]{Magrini:2022}, and several flags from Gaia~DR3 suggest binarity for KIC\,6634419 (e.g., RUWE = 9.972, \texttt{ipd\_frac\_multi\_peak} = 99). A large [Ba/Fe] enhancement combined with binarity may suggest that this object could be a Ba star, although none of the other s-process elements follow Ba.

KIC\,6605673 was included in the program because its [Fe/H] determination was flagged as uncertain in \citet[][]{Casagrande:2014aa}. Initially classified as a low-mass, metal-poor object, in our analysis it appears to be a regular thin disc dwarf star with near-solar metallicity and 1.08 $\pm$ 0.03 M$_{\odot}$. Meanwhile, KIC\,6940126, already discussed in Section~\ref{sec:resultages}, still has uncertain estimations even in its chemistry. As previously mentioned, different solutions in grid-based modelling may yield surface gravity values sufficiently different to impact the abundance ratios and $\alpha$-classification. Hence, the application of our method to KIC\,6940126 requires short-cadence observations to expand the range of frequencies available in its power spectrum.

From the point of view of chemical abundances, KIC\,8145677 has a [Fe/H] similar to the globular cluster 47~Tuc, \citep[][]{Carretta:2004,Alves-Brito:2005aa}, and it fits well in the red clump of that globular cluster in a Kiel diagram. However, from its kinematics we can rule out membership, as it has a prograde and eccentric orbit, while 47~Tuc is retrograde \citep[][]{Gaia:2018}. It is an old member of the thick disc, displaying enhancement in $\alpha$-elements, low [Mn/Fe], and being at the same time deficient in Eu while enhanced in s-process elements. Regarding its mass, given that it is a low-mass RC star, all the caveats discussed in Section~\ref{sec:omr} regarding uncertainties in the RGB mass loss apply here as well.

\section{Conclusions}
\label{sec:conclusions}

In this work we performed a detailed chrono-chemo-dynamical study of a sample of 16 stars, 15 of them being red giants, with light curves from \emph{Kepler} available. The sample selection focused on stars with perceived disagreements between their chemistry and their ages in previous works, and we followed a similar methodological prescription as in \citet[][]{AlencastroPuls:2022}. Here is a summary of our main findings.

\begin{itemize}
    
    \item The panels shown in Fig.~\ref{fig:mass_unc_vs_numax_dnu_unc} suggest that, for stellar masses, our method is robust and yields good overall precision. Targets with larger error bars in mass have had some problem in the extraction of $\Delta \nu$ or $\nu_{\mathrm{max}}$, and precision levels published here tend to be higher than those from large surveys, when the same level of precision in the observational inputs is considered.
    
    \item The two stars analysed in Section~\ref{sec:omr}, KIC\,6936796 and KIC\,7595155 appear to be regular thin disc metal-rich stars, except for their very old ages. One possibility is that they originated in the inner Galaxy but experienced radial migration. On the other hand, the two are low-mass RC stars, and their age determination might be inaccurate due to uncertainties in mass loss along the RGB.
    
    \item Regarding the young $\alpha$-rich stars in common with \citet[][]{Martig:2015aa} and \citet[][]{Hekker:2019}, we found some results conflicting with those previously reported in the literature. In particular, from our results only the data for KIC\,10525475 support the interpretation that this star is in fact an older giant which underwent mass accretion thus returning misleading young ages. Also, KIC\,3833399 seems to be $\alpha$-normal, in agreement with \citeauthor{Hekker:2019}. For the other four objects whose ages seem to be young, the results are ambiguous: their ages, kinematics and chemistry do not agree altogether.
    
    \item Upper-RGB stars have had problematic determination of ages also in the more-metal rich regime \citep[see][for discussion on metal-poor giants]{AlencastroPuls:2022}. This is particularly true for KIC\,9002884, whose age posterior in BASTA is double-peaked. While its adopted age is 13.1~Gyr (median of the distribution), our analysis identified a secondary peak around 8~Gyr. Hence, we suggest that future analysis like the one performed in this work focus on stars whose $\nu_{\mathrm{max}}$ is higher than 10~$\mu$Hz, unless probing into the upper-RGB is required.
    
    \item We have found some strong temporal correlations in some [X/Y] abundance ratios, in particular for Zn. While there is no evidence so far that the correlations found here have any physical significance, due to the particularities of the sample, we suggest that future studies on asteroseismic chemical clocks in red giants should pay attention to Zn.
    
\end{itemize}

Summing up, this paper clearly demonstrates that very high precision in asteroseismic masses and ages can be achieved with grid-based modelling for red giant stars whose $\nu_{\mathrm{max}}$ is higher than 10~$\mu$Hz, even if parallaxes are discarded. For RC stars, the accuracy in masses, hence ages, depends on the RGB mass loss. A key ingredient in this paper is \textsc{basta}, which interpolates all the available observables in a grid of models, and makes such high precision achievable as long as high quality measurements are available. The combination of ages, chemical abundances and dynamics is important to get the full picture of the evolution of the Galaxy in a more direct manner, and clearly there are outliers in the age-metallicity space which challenge our understanding of stellar nucleosyntheis, chemical evolution and/or the dynamical evolution of our Galaxy. A better understanding of mass loss in the RGB is urgently needed to clarify the picture in the lower mass end of the RC.

\section*{Acknowledgements}

We thank the anonymous referee for the comments and suggestions. Parts of this research were conducted by the Australian Research Council Centre of Excellence for All Sky Astrophysics in 3 Dimensions (ASTRO~3D), through project number CE170100013.
AAP acknowledges the support by the State of Hesse within the Research Cluster ELEMENTS (Project ID 500/10.006).
Funding for the Stellar Astrophysics Centre is provided by The Danish National Research Foundation (Grant agreement no.: DNRF106).
The spectroscopic data presented herein were obtained at the W. M. Keck Observatory, which is operated as a scientific partnership among the California Institute of Technology, the University of California and the National Aeronautics and Space Administration. The Observatory was made possible by the generous financial support of the W. M. Keck Foundation. The authors wish to recognize and acknowledge the very significant cultural role and reverence that the summit of Maunakea has always had within the indigenous Hawaiian community. We are most fortunate to have the opportunity to conduct observations from this mountain.
This research made use of \textsc{astropy}, \url{http://www.astropy.org}, a community-developed core \textsc{pyhton} package for Astronomy \citep{astropy:2013,astropy:2018}, \textsc{numpy} \citep{numpy:2020}, \textsc{matplotlib} \citep{Matplotlib:2007}, \textsc{scipy} \citep{scipy:2020}, and \textsc{Lightkurve}, a \textsc{python} package for Kepler and TESS data analysis \citep{lightkurve:2018}.
This publication makes use of data products from the Two Micron All Sky Survey, which is a joint project of the University of Massachusetts and the Infrared Processing and Analysis Center/California Institute of Technology, funded by the National Aeronautics and Space Administration and the National Science Foundation.

\section*{Data Availability}

The data underlying this article are available in the article and online through supplementary material.



\bibliographystyle{mnras}
\typeout{}
\bibliography{references} 

\begin{thebibliography}{}
\makeatletter
\relax
\def\mn@urlcharsother{\let\do\@makeother \do\$\do\&\do\#\do\^\do\_\do\%\do\~}
\def\mn@doi{\begingroup\mn@urlcharsother \@ifnextchar [ {\mn@doi@}
  {\mn@doi@[]}}
\def\mn@doi@[#1]#2{\def\@tempa{#1}\ifx\@tempa\@empty \href
  {http://dx.doi.org/#2} {doi:#2}\else \href {http://dx.doi.org/#2} {#1}\fi
  \endgroup}
\def\mn@eprint#1#2{\mn@eprint@#1:#2::\@nil}
\def\mn@eprint@arXiv#1{\href {http://arxiv.org/abs/#1} {{\tt arXiv:#1}}}
\def\mn@eprint@dblp#1{\href {http://dblp.uni-trier.de/rec/bibtex/#1.xml}
  {dblp:#1}}
\def\mn@eprint@#1:#2:#3:#4\@nil{\def\@tempa {#1}\def\@tempb {#2}\def\@tempc
  {#3}\ifx \@tempc \@empty \let \@tempc \@tempb \let \@tempb \@tempa \fi \ifx
  \@tempb \@empty \def\@tempb {arXiv}\fi \@ifundefined
  {mn@eprint@\@tempb}{\@tempb:\@tempc}{\expandafter \expandafter \csname
  mn@eprint@\@tempb\endcsname \expandafter{\@tempc}}}

\bibitem[\protect\citeauthoryear{{Aguirre B{\o}rsen-Koch} et~al.,}{{Aguirre
  B{\o}rsen-Koch} et~al.}{2022}]{Aguirre:2022}
{Aguirre B{\o}rsen-Koch} V.,  et~al., 2022, \mn@doi [\mnras]
  {10.1093/mnras/stab2911}, \href
  {https://ui.adsabs.harvard.edu/abs/2022MNRAS.509.4344A} {509, 4344}

\bibitem[\protect\citeauthoryear{{Alencastro Puls}, {Casagrande}, {Monty},
  {Yong}, {Liu}, {Stello}, {Aguirre B{\o}rsen-Koch}  \& {Freeman}}{{Alencastro
  Puls} et~al.}{2022}]{AlencastroPuls:2022}
{Alencastro Puls} A.,  {Casagrande} L.,  {Monty} S.,  {Yong} D.,  {Liu} F.,
  {Stello} D.,  {Aguirre B{\o}rsen-Koch} V.,   {Freeman} K.~C.,  2022, \mn@doi
  [\mnras] {10.1093/mnras/stab3545}, \href
  {https://ui.adsabs.harvard.edu/abs/2022MNRAS.510.1733A} {510, 1733}

\bibitem[\protect\citeauthoryear{{Alves-Brito} et~al.,}{{Alves-Brito}
  et~al.}{2005}]{Alves-Brito:2005aa}
{Alves-Brito} A.,  et~al., 2005, \mn@doi [\aap] {10.1051/0004-6361:20041634},
  \href {http://adsabs.harvard.edu/abs/2005A%26A...435..657A} {435, 657}

\bibitem[\protect\citeauthoryear{{Alves-Brito}, {Mel{\'e}ndez}, {Asplund},
  {Ram{\'{\i}}rez}  \& {Yong}}{{Alves-Brito} et~al.}{2010}]{Alves-Brito:2010aa}
{Alves-Brito} A.,  {Mel{\'e}ndez} J.,  {Asplund} M.,  {Ram{\'{\i}}rez} I.,
  {Yong} D.,  2010, \mn@doi [\aap] {10.1051/0004-6361/200913444}, \href
  {http://adsabs.harvard.edu/abs/2010A%26A...513A..35A} {513, A35}

\bibitem[\protect\citeauthoryear{{Asplund}, {Grevesse}, {Sauval}  \&
  {Scott}}{{Asplund} et~al.}{2009}]{Asplund:2009aa}
{Asplund} M.,  {Grevesse} N.,  {Sauval} A.~J.,   {Scott} P.,  2009, \mn@doi
  [\araa] {10.1146/annurev.astro.46.060407.145222}, \href
  {http://adsabs.harvard.edu/abs/2009ARA%26A..47..481A} {47, 481}

\bibitem[\protect\citeauthoryear{{Astropy Collaboration} et~al.,}{{Astropy
  Collaboration} et~al.}{2013}]{astropy:2013}
{Astropy Collaboration} et~al., 2013, \mn@doi [\aap]
  {10.1051/0004-6361/201322068}, \href
  {http://adsabs.harvard.edu/abs/2013A%26A...558A..33A} {558, A33}

\bibitem[\protect\citeauthoryear{{Astropy Collaboration} et~al.,}{{Astropy
  Collaboration} et~al.}{2018}]{astropy:2018}
{Astropy Collaboration} et~al., 2018, \mn@doi [\aj] {10.3847/1538-3881/aabc4f},
  \href {https://ui.adsabs.harvard.edu/abs/2018AJ....156..123A} {156, 123}

\bibitem[\protect\citeauthoryear{{Baglin} et~al.,}{{Baglin}
  et~al.}{2006}]{Baglin:2006aa}
{Baglin} A.,  et~al., 2006, in 36th COSPAR Scientific Assembly. p.~3749

\bibitem[\protect\citeauthoryear{{Bailer-Jones}, {Rybizki}, {Fouesneau},
  {Mantelet}  \& {Andrae}}{{Bailer-Jones} et~al.}{2018}]{BailerJones:2018}
{Bailer-Jones} C.~A.~L.,  {Rybizki} J.,  {Fouesneau} M.,  {Mantelet} G.,
  {Andrae} R.,  2018, \mn@doi [\aj] {10.3847/1538-3881/aacb21}, \href
  {https://ui.adsabs.harvard.edu/abs/2018AJ....156...58B} {156, 58}

\bibitem[\protect\citeauthoryear{{Bailer-Jones}, {Rybizki}, {Fouesneau},
  {Demleitner}  \& {Andrae}}{{Bailer-Jones} et~al.}{2021}]{BailerJones:2021}
{Bailer-Jones} C.~A.~L.,  {Rybizki} J.,  {Fouesneau} M.,  {Demleitner} M.,
  {Andrae} R.,  2021, \mn@doi [\aj] {10.3847/1538-3881/abd806}, \href
  {https://ui.adsabs.harvard.edu/abs/2021AJ....161..147B} {161, 147}

\bibitem[\protect\citeauthoryear{{Barbuy} et~al.,}{{Barbuy}
  et~al.}{2013}]{Barbuy:2013}
{Barbuy} B.,  et~al., 2013, \mn@doi [\aap] {10.1051/0004-6361/201322380}, \href
  {https://ui.adsabs.harvard.edu/abs/2013A&A...559A...5B} {559, A5}

\bibitem[\protect\citeauthoryear{{Bensby}, {Alves-Brito}, {Oey}, {Yong}  \&
  {Mel{\'e}ndez}}{{Bensby} et~al.}{2011}]{Bensby:2011aa}
{Bensby} T.,  {Alves-Brito} A.,  {Oey} M.~S.,  {Yong} D.,   {Mel{\'e}ndez} J.,
  2011, \mn@doi [\apjl] {10.1088/2041-8205/735/2/L46}, \href
  {http://adsabs.harvard.edu/abs/2011ApJ...735L..46B} {735, L46}

\bibitem[\protect\citeauthoryear{{Bergemann}}{{Bergemann}}{2011}]{Bergemann:2011aa}
{Bergemann} M.,  2011, \mn@doi [\mnras] {10.1111/j.1365-2966.2011.18295.x},
  \href {http://adsabs.harvard.edu/abs/2011MNRAS.413.2184B} {413, 2184}

\bibitem[\protect\citeauthoryear{{Bergemann} \& {Cescutti}}{{Bergemann} \&
  {Cescutti}}{2010}]{Bergemann:2010b}
{Bergemann} M.,  {Cescutti} G.,  2010, \mn@doi [\aap]
  {10.1051/0004-6361/201014250}, \href
  {https://ui.adsabs.harvard.edu/abs/2010A&A...522A...9B} {522, A9}

\bibitem[\protect\citeauthoryear{{Bergemann}, {Pickering}  \&
  {Gehren}}{{Bergemann} et~al.}{2010}]{Bergemann:2010}
{Bergemann} M.,  {Pickering} J.~C.,   {Gehren} T.,  2010, \mn@doi [\mnras]
  {10.1111/j.1365-2966.2009.15736.x}, \href
  {https://ui.adsabs.harvard.edu/abs/2010MNRAS.401.1334B} {401, 1334}

\bibitem[\protect\citeauthoryear{{Bergemann}, {Lind}, {Collet}, {Magic}  \&
  {Asplund}}{{Bergemann} et~al.}{2012}]{Bergemann:2012aa}
{Bergemann} M.,  {Lind} K.,  {Collet} R.,  {Magic} Z.,   {Asplund} M.,  2012,
  \mn@doi [\mnras] {10.1111/j.1365-2966.2012.21687.x}, \href
  {http://adsabs.harvard.edu/abs/2012MNRAS.427...27B} {427, 27}

\bibitem[\protect\citeauthoryear{{Bergemann}, {Kudritzki}, {W{\"u}rl}, {Plez},
  {Davies}  \& {Gazak}}{{Bergemann} et~al.}{2013}]{Bergemann:2013}
{Bergemann} M.,  {Kudritzki} R.-P.,  {W{\"u}rl} M.,  {Plez} B.,  {Davies} B.,
  {Gazak} Z.,  2013, \mn@doi [\apj] {10.1088/0004-637X/764/2/115}, \href
  {https://ui.adsabs.harvard.edu/abs/2013ApJ...764..115B} {764, 115}

\bibitem[\protect\citeauthoryear{{Bergemann} et~al.,}{{Bergemann}
  et~al.}{2019}]{Bergemann:2019}
{Bergemann} M.,  et~al., 2019, \mn@doi [\aap] {10.1051/0004-6361/201935811},
  \href {https://ui.adsabs.harvard.edu/abs/2019A&A...631A..80B} {631, A80}

\bibitem[\protect\citeauthoryear{{Bergemann} et~al.,}{{Bergemann}
  et~al.}{2021}]{Bergemann:2021}
{Bergemann} M.,  et~al., 2021, \mn@doi [\mnras] {10.1093/mnras/stab2160}, \href
  {https://ui.adsabs.harvard.edu/abs/2021MNRAS.508.2236B} {508, 2236}

\bibitem[\protect\citeauthoryear{{Binney}}{{Binney}}{2012}]{Binney:2012}
{Binney} J.,  2012, \mn@doi [\mnras] {10.1111/j.1365-2966.2012.21757.x}, \href
  {https://ui.adsabs.harvard.edu/abs/2012MNRAS.426.1324B} {426, 1324}

\bibitem[\protect\citeauthoryear{{Blackwell} \& {Shallis}}{{Blackwell} \&
  {Shallis}}{1977}]{Blackwell:1977aa}
{Blackwell} D.~E.,  {Shallis} M.~J.,  1977, \mn@doi [\mnras]
  {10.1093/mnras/180.2.177}, \href
  {http://adsabs.harvard.edu/abs/1977MNRAS.180..177B} {180, 177}

\bibitem[\protect\citeauthoryear{{Bovy}}{{Bovy}}{2015}]{Bovy:2015}
{Bovy} J.,  2015, \mn@doi [\apjs] {10.1088/0067-0049/216/2/29}, \href
  {https://ui.adsabs.harvard.edu/abs/2015ApJS..216...29B} {216, 29}

\bibitem[\protect\citeauthoryear{{Bressan}, {Marigo}, {Girardi}, {Salasnich},
  {Dal Cero}, {Rubele}  \& {Nanni}}{{Bressan} et~al.}{2012}]{Bressan:2012aa}
{Bressan} A.,  {Marigo} P.,  {Girardi} L.,  {Salasnich} B.,  {Dal Cero} C.,
  {Rubele} S.,   {Nanni} A.,  2012, \mn@doi [\mnras]
  {10.1111/j.1365-2966.2012.21948.x}, \href
  {http://adsabs.harvard.edu/abs/2012MNRAS.427..127B} {427, 127}

\bibitem[\protect\citeauthoryear{{Buder} et~al.,}{{Buder}
  et~al.}{2018}]{Buder:2018aa}
{Buder} S.,  et~al., 2018, \mn@doi [\mnras] {10.1093/mnras/sty1281}, \href
  {http://adsabs.harvard.edu/abs/2018MNRAS.478.4513B} {478, 4513}

\bibitem[\protect\citeauthoryear{{Buder} et~al.,}{{Buder}
  et~al.}{2021}]{Buder:2021}
{Buder} S.,  et~al., 2021, \mn@doi [\mnras] {10.1093/mnras/stab1242}, \href
  {https://ui.adsabs.harvard.edu/abs/2021MNRAS.506..150B} {506, 150}

\bibitem[\protect\citeauthoryear{{Burbidge}, {Burbidge}, {Fowler}  \&
  {Hoyle}}{{Burbidge} et~al.}{1957}]{Burbidge:1957aa}
{Burbidge} E.~M.,  {Burbidge} G.~R.,  {Fowler} W.~A.,   {Hoyle} F.,  1957,
  \mn@doi [Reviews of Modern Physics] {10.1103/RevModPhys.29.547}, \href
  {http://adsabs.harvard.edu/abs/1957RvMP...29..547B} {29, 547}

\bibitem[\protect\citeauthoryear{{Carretta}, {Gratton}, {Bragaglia},
  {Bonifacio}  \& {Pasquini}}{{Carretta} et~al.}{2004}]{Carretta:2004}
{Carretta} E.,  {Gratton} R.~G.,  {Bragaglia} A.,  {Bonifacio} P.,   {Pasquini}
  L.,  2004, \mn@doi [\aap] {10.1051/0004-6361:20034370}, \href
  {https://ui.adsabs.harvard.edu/abs/2004A&A...416..925C} {416, 925}

\bibitem[\protect\citeauthoryear{{Casagrande}, {Ram{\'{\i}}rez},
  {Mel{\'e}ndez}, {Bessell}  \& {Asplund}}{{Casagrande}
  et~al.}{2010}]{Casagrande:2010aa}
{Casagrande} L.,  {Ram{\'{\i}}rez} I.,  {Mel{\'e}ndez} J.,  {Bessell} M.,
  {Asplund} M.,  2010, \mn@doi [\aap] {10.1051/0004-6361/200913204}, \href
  {http://adsabs.harvard.edu/abs/2010A%26A...512A..54C} {512, A54}

\bibitem[\protect\citeauthoryear{{Casagrande}, {Sch{\"o}nrich}, {Asplund},
  {Cassisi}, {Ram{\'{\i}}rez}, {Mel{\'e}ndez}, {Bensby}  \&
  {Feltzing}}{{Casagrande} et~al.}{2011}]{Casagrande:2011aa}
{Casagrande} L.,  {Sch{\"o}nrich} R.,  {Asplund} M.,  {Cassisi} S.,
  {Ram{\'{\i}}rez} I.,  {Mel{\'e}ndez} J.,  {Bensby} T.,   {Feltzing} S.,
  2011, \mn@doi [\aap] {10.1051/0004-6361/201016276}, \href
  {http://adsabs.harvard.edu/abs/2011A%26A...530A.138C} {530, A138}

\bibitem[\protect\citeauthoryear{{Casagrande} et~al.,}{{Casagrande}
  et~al.}{2014}]{Casagrande:2014aa}
{Casagrande} L.,  et~al., 2014, \mn@doi [\apj] {10.1088/0004-637X/787/2/110},
  \href {http://adsabs.harvard.edu/abs/2014ApJ...787..110C} {787, 110}

\bibitem[\protect\citeauthoryear{{Casagrande} et~al.,}{{Casagrande}
  et~al.}{2016}]{Casagrande:2016aa}
{Casagrande} L.,  et~al., 2016, \mn@doi [\mnras] {10.1093/mnras/stv2320}, \href
  {http://adsabs.harvard.edu/abs/2016MNRAS.455..987C} {455, 987}

\bibitem[\protect\citeauthoryear{{Casagrande} et~al.,}{{Casagrande}
  et~al.}{2021}]{Casagrande:2021}
{Casagrande} L.,  et~al., 2021, \mn@doi [\mnras] {10.1093/mnras/stab2304},
  \href {https://ui.adsabs.harvard.edu/abs/2021MNRAS.507.2684C} {507, 2684}

\bibitem[\protect\citeauthoryear{{Castelli} \& {Kurucz}}{{Castelli} \&
  {Kurucz}}{2003}]{Castelli:2003aa}
{Castelli} F.,  {Kurucz} R.~L.,  2003, in {Piskunov} N.,  {Weiss} W.~W.,
  {Gray} D.~F.,  eds,  IAU Symposium Vol. 210, Modelling of Stellar
  Atmospheres. p.~20P

\bibitem[\protect\citeauthoryear{{Cayrel} et~al.,}{{Cayrel}
  et~al.}{2004}]{Cayrel:2004aa}
{Cayrel} R.,  et~al., 2004, \mn@doi [\aap] {10.1051/0004-6361:20034074}, \href
  {http://adsabs.harvard.edu/abs/2004A%26A...416.1117C} {416, 1117}

\bibitem[\protect\citeauthoryear{{Chiappini} et~al.,}{{Chiappini}
  et~al.}{2015}]{Chiappini:2015aa}
{Chiappini} C.,  et~al., 2015, \mn@doi [\aap] {10.1051/0004-6361/201525865},
  \href {http://adsabs.harvard.edu/abs/2015A%26A...576L..12C} {576, L12}

\bibitem[\protect\citeauthoryear{{Cutri} et~al.,}{{Cutri}
  et~al.}{2003}]{Cutri:2003}
{Cutri} R.~M.,  et~al., 2003, {2MASS All Sky Catalog of point sources.}

\bibitem[\protect\citeauthoryear{{Dotter}, {Conroy}, {Cargile}  \&
  {Asplund}}{{Dotter} et~al.}{2017}]{Dotter:2017aa}
{Dotter} A.,  {Conroy} C.,  {Cargile} P.,   {Asplund} M.,  2017, \mn@doi [\apj]
  {10.3847/1538-4357/aa6d10}, \href
  {http://adsabs.harvard.edu/abs/2017ApJ...840...99D} {840, 99}

\bibitem[\protect\citeauthoryear{{Edvardsson}, {Andersen}, {Gustafsson},
  {Lambert}, {Nissen}  \& {Tomkin}}{{Edvardsson}
  et~al.}{1993}]{Edvardsson:1993aa}
{Edvardsson} B.,  {Andersen} J.,  {Gustafsson} B.,  {Lambert} D.~L.,  {Nissen}
  P.~E.,   {Tomkin} J.,  1993, \aap, \href
  {http://adsabs.harvard.edu/abs/1993A%26A...275..101E} {275, 101}

\bibitem[\protect\citeauthoryear{{Epstein}, {Johnson}, {Dong}, {Udalski},
  {Gould}  \& {Becker}}{{Epstein} et~al.}{2010}]{Epstein:2010aa}
{Epstein} C.~R.,  {Johnson} J.~A.,  {Dong} S.,  {Udalski} A.,  {Gould} A.,
  {Becker} G.,  2010, \mn@doi [\apj] {10.1088/0004-637X/709/1/447}, \href
  {http://adsabs.harvard.edu/abs/2010ApJ...709..447E} {709, 447}

\bibitem[\protect\citeauthoryear{{Freeman} \& {Bland-Hawthorn}}{{Freeman} \&
  {Bland-Hawthorn}}{2002}]{Freeman:2002aa}
{Freeman} K.,  {Bland-Hawthorn} J.,  2002, \mn@doi [\araa]
  {10.1146/annurev.astro.40.060401.093840}, \href
  {http://adsabs.harvard.edu/abs/2002ARA%26A..40..487F} {40, 487}

\bibitem[\protect\citeauthoryear{{Gaia Collaboration} et~al.,}{{Gaia
  Collaboration} et~al.}{2018a}]{GaiaDR2:2018}
{Gaia Collaboration} et~al., 2018a, \mn@doi [\aap]
  {10.1051/0004-6361/201833051}, \href
  {https://ui.adsabs.harvard.edu/abs/2018A&A...616A...1G} {616, A1}

\bibitem[\protect\citeauthoryear{{Gaia Collaboration} et~al.,}{{Gaia
  Collaboration} et~al.}{2018b}]{Gaia:2018}
{Gaia Collaboration} et~al., 2018b, \mn@doi [\aap]
  {10.1051/0004-6361/201832698}, \href
  {https://ui.adsabs.harvard.edu/abs/2018A&A...616A..12G} {616, A12}

\bibitem[\protect\citeauthoryear{{Gaia Collaboration} et~al.,}{{Gaia
  Collaboration} et~al.}{2021}]{GaiaEDR3:2021}
{Gaia Collaboration} et~al., 2021, \mn@doi [\aap]
  {10.1051/0004-6361/202039657}, \href
  {https://ui.adsabs.harvard.edu/abs/2021A&A...649A...1G} {649, A1}

\bibitem[\protect\citeauthoryear{{Gaia Collaboration} et~al.,}{{Gaia
  Collaboration} et~al.}{2022}]{2022arXiv220800211G}
{Gaia Collaboration} et~al., 2022, arXiv e-prints, \href
  {https://ui.adsabs.harvard.edu/abs/2022arXiv220800211G} {p. arXiv:2208.00211}

\bibitem[\protect\citeauthoryear{{Green}, {Schlafly}, {Zucker}, {Speagle}  \&
  {Finkbeiner}}{{Green} et~al.}{2019}]{Green:2019}
{Green} G.~M.,  {Schlafly} E.,  {Zucker} C.,  {Speagle} J.~S.,   {Finkbeiner}
  D.,  2019, \mn@doi [\apj] {10.3847/1538-4357/ab5362}, \href
  {https://ui.adsabs.harvard.edu/abs/2019ApJ...887...93G} {887, 93}

\bibitem[\protect\citeauthoryear{{Handberg}, {Brogaard}, {Miglio}, {Bossini},
  {Elsworth}, {Slumstrup}, {Davies}  \& {Chaplin}}{{Handberg}
  et~al.}{2017}]{Handberg:2017}
{Handberg} R.,  {Brogaard} K.,  {Miglio} A.,  {Bossini} D.,  {Elsworth} Y.,
  {Slumstrup} D.,  {Davies} G.~R.,   {Chaplin} W.~J.,  2017, \mn@doi [\mnras]
  {10.1093/mnras/stx1929}, \href
  {https://ui.adsabs.harvard.edu/abs/2017MNRAS.472..979H} {472, 979}

\bibitem[\protect\citeauthoryear{Harris et~al.,}{Harris
  et~al.}{2020}]{numpy:2020}
Harris C.~R.,  et~al., 2020, \mn@doi [Nature] {10.1038/s41586-020-2649-2}, 585,
  357

\bibitem[\protect\citeauthoryear{{Harvey}}{{Harvey}}{1985}]{Harvey:1985}
{Harvey} J.,  1985, in {Rolfe} E.,  {Battrick} B.,  eds,  ESA Special
  Publication Vol. 235, Future Missions in Solar, Heliospheric \& Space Plasma
  Physics. p.~199

\bibitem[\protect\citeauthoryear{{Hawkins}, {Masseron}, {Jofr{\'e}}, {Gilmore},
  {Elsworth}  \& {Hekker}}{{Hawkins} et~al.}{2016}]{Hawkins:2016}
{Hawkins} K.,  {Masseron} T.,  {Jofr{\'e}} P.,  {Gilmore} G.,  {Elsworth} Y.,
  {Hekker} S.,  2016, \mn@doi [\aap] {10.1051/0004-6361/201628812}, \href
  {https://ui.adsabs.harvard.edu/abs/2016A&A...594A..43H} {594, A43}

\bibitem[\protect\citeauthoryear{{Haywood}}{{Haywood}}{2008}]{Haywood:2008}
{Haywood} M.,  2008, \mn@doi [\mnras] {10.1111/j.1365-2966.2008.13395.x}, \href
  {https://ui.adsabs.harvard.edu/abs/2008MNRAS.388.1175H} {388, 1175}

\bibitem[\protect\citeauthoryear{{Hekker} \& {Johnson}}{{Hekker} \&
  {Johnson}}{2019}]{Hekker:2019}
{Hekker} S.,  {Johnson} J.~A.,  2019, \mn@doi [\mnras] {10.1093/mnras/stz1554},
  \href {https://ui.adsabs.harvard.edu/abs/2019MNRAS.487.4343H} {487, 4343}

\bibitem[\protect\citeauthoryear{{Hidalgo} et~al.,}{{Hidalgo}
  et~al.}{2018}]{Hidalgo:2018}
{Hidalgo} S.~L.,  et~al., 2018, \mn@doi [\apj] {10.3847/1538-4357/aab158},
  \href {https://ui.adsabs.harvard.edu/abs/2018ApJ...856..125H} {856, 125}

\bibitem[\protect\citeauthoryear{{Huber} et~al.,}{{Huber}
  et~al.}{2011}]{Huber:2011b}
{Huber} D.,  et~al., 2011, \mn@doi [\apj] {10.1088/0004-637X/743/2/143}, \href
  {https://ui.adsabs.harvard.edu/abs/2011ApJ...743..143H} {743, 143}

\bibitem[\protect\citeauthoryear{Hunter}{Hunter}{2007}]{Matplotlib:2007}
Hunter J.~D.,  2007, \mn@doi [Computing in Science \& Engineering]
  {10.1109/MCSE.2007.55}, 9, 90

\bibitem[\protect\citeauthoryear{{Ishigaki}, {Aoki}  \& {Chiba}}{{Ishigaki}
  et~al.}{2013}]{Ishigaki:2013}
{Ishigaki} M.~N.,  {Aoki} W.,   {Chiba} M.,  2013, \mn@doi [\apj]
  {10.1088/0004-637X/771/1/67}, \href
  {https://ui.adsabs.harvard.edu/abs/2013ApJ...771...67I} {771, 67}

\bibitem[\protect\citeauthoryear{{Jofr{\'e}} et~al.,}{{Jofr{\'e}}
  et~al.}{2016}]{Jofre:2016aa}
{Jofr{\'e}} P.,  et~al., 2016, \mn@doi [\aap] {10.1051/0004-6361/201629356},
  \href {http://adsabs.harvard.edu/abs/2016A%26A...595A..60J} {595, A60}

\bibitem[\protect\citeauthoryear{{Jofr{\'e}} et~al.,}{{Jofr{\'e}}
  et~al.}{2023}]{Jofre:2023}
{Jofr{\'e}} P.,  et~al., 2023, \mn@doi [\aap] {10.1051/0004-6361/202244524},
  \href {https://ui.adsabs.harvard.edu/abs/2023A&A...671A..21J} {671, A21}

\bibitem[\protect\citeauthoryear{{J{\"o}nsson} et~al.,}{{J{\"o}nsson}
  et~al.}{2020}]{Jonsson:2020}
{J{\"o}nsson} H.,  et~al., 2020, \mn@doi [\aj] {10.3847/1538-3881/aba592},
  \href {https://ui.adsabs.harvard.edu/abs/2020AJ....160..120J} {160, 120}

\bibitem[\protect\citeauthoryear{{Karakas} \& {Lugaro}}{{Karakas} \&
  {Lugaro}}{2016}]{Karakas:2016aa}
{Karakas} A.~I.,  {Lugaro} M.,  2016, \mn@doi [\apj]
  {10.3847/0004-637X/825/1/26}, \href
  {http://adsabs.harvard.edu/abs/2016ApJ...825...26K} {825, 26}

\bibitem[\protect\citeauthoryear{{Karakas}, {Marino}  \& {Nataf}}{{Karakas}
  et~al.}{2014}]{Karakas:2014aa}
{Karakas} A.~I.,  {Marino} A.~F.,   {Nataf} D.~M.,  2014, \mn@doi [\apj]
  {10.1088/0004-637X/784/1/32}, \href
  {http://adsabs.harvard.edu/abs/2014ApJ...784...32K} {784, 32}

\bibitem[\protect\citeauthoryear{{Karoff}}{{Karoff}}{2012}]{Karoff:2012}
{Karoff} C.,  2012, \mn@doi [\mnras] {10.1111/j.1365-2966.2012.20542.x}, \href
  {https://ui.adsabs.harvard.edu/abs/2012MNRAS.421.3170K} {421, 3170}

\bibitem[\protect\citeauthoryear{{Kilic}, {Munn}, {Harris}, {von Hippel},
  {Liebert}, {Williams}, {Jeffery}  \& {DeGennaro}}{{Kilic}
  et~al.}{2017}]{Kilic:2017}
{Kilic} M.,  {Munn} J.~A.,  {Harris} H.~C.,  {von Hippel} T.,  {Liebert} J.~W.,
   {Williams} K.~A.,  {Jeffery} E.,   {DeGennaro} S.,  2017, \mn@doi [\apj]
  {10.3847/1538-4357/aa62a5}, \href
  {https://ui.adsabs.harvard.edu/abs/2017ApJ...837..162K} {837, 162}

\bibitem[\protect\citeauthoryear{{Kobayashi}, {Karakas}  \&
  {Lugaro}}{{Kobayashi} et~al.}{2020}]{Kobayashi:2020}
{Kobayashi} C.,  {Karakas} A.~I.,   {Lugaro} M.,  2020, \mn@doi [\apj]
  {10.3847/1538-4357/abae65}, \href
  {https://ui.adsabs.harvard.edu/abs/2020ApJ...900..179K} {900, 179}

\bibitem[\protect\citeauthoryear{{Koch} et~al.,}{{Koch}
  et~al.}{2010}]{Koch:2010}
{Koch} D.~G.,  et~al., 2010, \mn@doi [\apjl] {10.1088/2041-8205/713/2/L79},
  \href {https://ui.adsabs.harvard.edu/abs/2010ApJ...713L..79K} {713, L79}

\bibitem[\protect\citeauthoryear{{Li} et~al.,}{{Li} et~al.}{2022}]{Li:2022}
{Li} Y.,  et~al., 2022, \mn@doi [Nature Astronomy]
  {10.1038/s41550-022-01648-5}, \href
  {https://ui.adsabs.harvard.edu/abs/2022NatAs...6..673L} {6, 673}

\bibitem[\protect\citeauthoryear{{Lightkurve Collaboration}
  et~al.,}{{Lightkurve Collaboration} et~al.}{2018}]{lightkurve:2018}
{Lightkurve Collaboration} et~al., 2018, {Lightkurve: Kepler and TESS time
  series analysis in Python}, Astrophysics Source Code Library (\mn@eprint
  {ascl} {1812.013})

\bibitem[\protect\citeauthoryear{{Lin} et~al.,}{{Lin} et~al.}{2020}]{Lin:2020}
{Lin} J.,  et~al., 2020, \mn@doi [\mnras] {10.1093/mnras/stz3048}, \href
  {https://ui.adsabs.harvard.edu/abs/2020MNRAS.491.2043L} {491, 2043}

\bibitem[\protect\citeauthoryear{{Lind}, {Asplund}, {Barklem}  \&
  {Belyaev}}{{Lind} et~al.}{2011}]{Lind:2011}
{Lind} K.,  {Asplund} M.,  {Barklem} P.~S.,   {Belyaev} A.~K.,  2011, \mn@doi
  [\aap] {10.1051/0004-6361/201016095}, \href
  {https://ui.adsabs.harvard.edu/abs/2011A&A...528A.103L} {528, A103}

\bibitem[\protect\citeauthoryear{{Lind}, {Bergemann}  \& {Asplund}}{{Lind}
  et~al.}{2012}]{Lind:2012aa}
{Lind} K.,  {Bergemann} M.,   {Asplund} M.,  2012, \mn@doi [\mnras]
  {10.1111/j.1365-2966.2012.21686.x}, \href
  {http://adsabs.harvard.edu/abs/2012MNRAS.427...50L} {427, 50}

\bibitem[\protect\citeauthoryear{{Lindegren} et~al.,}{{Lindegren}
  et~al.}{2021}]{Lindegren:2021}
{Lindegren} L.,  et~al., 2021, \mn@doi [\aap] {10.1051/0004-6361/202039653},
  \href {https://ui.adsabs.harvard.edu/abs/2021A&A...649A...4L} {649, A4}

\bibitem[\protect\citeauthoryear{{Lund} et~al.,}{{Lund}
  et~al.}{2016}]{Lund:2016}
{Lund} M.~N.,  et~al., 2016, \mn@doi [\pasp]
  {10.1088/1538-3873/128/970/124204}, \href
  {https://ui.adsabs.harvard.edu/abs/2016PASP..128l4204L} {128, 124204}

\bibitem[\protect\citeauthoryear{{Lund} et~al.,}{{Lund}
  et~al.}{2017}]{Lund:2017}
{Lund} M.~N.,  et~al., 2017, \mn@doi [\apj] {10.3847/1538-4357/835/2/172},
  \href {https://ui.adsabs.harvard.edu/abs/2017ApJ...835..172L} {835, 172}

\bibitem[\protect\citeauthoryear{{Mackereth} \& {Bovy}}{{Mackereth} \&
  {Bovy}}{2018}]{Mackereth:2018}
{Mackereth} J.~T.,  {Bovy} J.,  2018, \mn@doi [\pasp]
  {10.1088/1538-3873/aadcdd}, \href
  {https://ui.adsabs.harvard.edu/abs/2018PASP..130k4501M} {130, 114501}

\bibitem[\protect\citeauthoryear{{Magrini} et~al.,}{{Magrini}
  et~al.}{2022}]{Magrini:2022}
{Magrini} L.,  et~al., 2022, \mn@doi [Universe] {10.3390/universe8020064},
  \href {https://ui.adsabs.harvard.edu/abs/2022Univ....8...64M} {8, 64}

\bibitem[\protect\citeauthoryear{{Martig} et~al.,}{{Martig}
  et~al.}{2015}]{Martig:2015aa}
{Martig} M.,  et~al., 2015, \mn@doi [\mnras] {10.1093/mnras/stv1071}, \href
  {http://adsabs.harvard.edu/abs/2015MNRAS.451.2230M} {451, 2230}

\bibitem[\protect\citeauthoryear{{Martig} et~al.,}{{Martig}
  et~al.}{2016}]{Martig:2016aa}
{Martig} M.,  et~al., 2016, \mn@doi [\mnras] {10.1093/mnras/stv2830}, \href
  {http://adsabs.harvard.edu/abs/2016MNRAS.456.3655M} {456, 3655}

\bibitem[\protect\citeauthoryear{{Mashonkina}, {Korn}  \&
  {Przybilla}}{{Mashonkina} et~al.}{2007}]{Mashonkina:2007}
{Mashonkina} L.,  {Korn} A.~J.,   {Przybilla} N.,  2007, \mn@doi [\aap]
  {10.1051/0004-6361:20065999}, \href
  {https://ui.adsabs.harvard.edu/abs/2007A&A...461..261M} {461, 261}

\bibitem[\protect\citeauthoryear{{Matsuno}, {Yong}, {Aoki}  \&
  {Ishigaki}}{{Matsuno} et~al.}{2018}]{Matsuno:2018aa}
{Matsuno} T.,  {Yong} D.,  {Aoki} W.,   {Ishigaki} M.~N.,  2018, \mn@doi [\apj]
  {10.3847/1538-4357/aac019}, \href
  {http://adsabs.harvard.edu/abs/2018ApJ...860...49M} {860, 49}

\bibitem[\protect\citeauthoryear{{Matteucci}}{{Matteucci}}{2021}]{Matteucci:2021}
{Matteucci} F.,  2021, \mn@doi [\aapr] {10.1007/s00159-021-00133-8}, \href
  {https://ui.adsabs.harvard.edu/abs/2021A&ARv..29....5M} {29, 5}

\bibitem[\protect\citeauthoryear{{Matteucci} \& {Recchi}}{{Matteucci} \&
  {Recchi}}{2001}]{Matteucci:2001}
{Matteucci} F.,  {Recchi} S.,  2001, \mn@doi [\apj] {10.1086/322472}, \href
  {https://ui.adsabs.harvard.edu/abs/2001ApJ...558..351M} {558, 351}

\bibitem[\protect\citeauthoryear{{McMillan}}{{McMillan}}{2017}]{McMillan:2017}
{McMillan} P.~J.,  2017, \mn@doi [\mnras] {10.1093/mnras/stw2759}, \href
  {https://ui.adsabs.harvard.edu/abs/2017MNRAS.465...76M} {465, 76}

\bibitem[\protect\citeauthoryear{{Mel{\'e}ndez} et~al.,}{{Mel{\'e}ndez}
  et~al.}{2012}]{Melendez:2012}
{Mel{\'e}ndez} J.,  et~al., 2012, \mn@doi [\aap] {10.1051/0004-6361/201117222},
  \href {https://ui.adsabs.harvard.edu/abs/2012A&A...543A..29M} {543, A29}

\bibitem[\protect\citeauthoryear{{Miglio} et~al.,}{{Miglio}
  et~al.}{2012}]{Miglio:2012}
{Miglio} A.,  et~al., 2012, \mn@doi [\mnras]
  {10.1111/j.1365-2966.2011.19859.x}, \href
  {https://ui.adsabs.harvard.edu/abs/2012MNRAS.419.2077M} {419, 2077}

\bibitem[\protect\citeauthoryear{{Miglio} et~al.,}{{Miglio}
  et~al.}{2021}]{Miglio:2021}
{Miglio} A.,  et~al., 2021, \mn@doi [\aap] {10.1051/0004-6361/202038307}, \href
  {https://ui.adsabs.harvard.edu/abs/2021A&A...645A..85M} {645, A85}

\bibitem[\protect\citeauthoryear{{Mosser} et~al.,}{{Mosser}
  et~al.}{2014}]{Mosser:2014}
{Mosser} B.,  et~al., 2014, \mn@doi [\aap] {10.1051/0004-6361/201425039}, \href
  {https://ui.adsabs.harvard.edu/abs/2014A&A...572L...5M} {572, L5}

\bibitem[\protect\citeauthoryear{{Ness}, {Hogg}, {Rix}, {Ho}  \&
  {Zasowski}}{{Ness} et~al.}{2015}]{Ness:2015aa}
{Ness} M.,  {Hogg} D.~W.,  {Rix} H.-W.,  {Ho} A.~Y.~Q.,   {Zasowski} G.,  2015,
  \mn@doi [\apj] {10.1088/0004-637X/808/1/16}, \href
  {http://adsabs.harvard.edu/abs/2015ApJ...808...16N} {808, 16}

\bibitem[\protect\citeauthoryear{{Ness}, {Johnston}, {Blancato}, {Rix},
  {Beane}, {Bird}  \& {Hawkins}}{{Ness} et~al.}{2019}]{Ness:2019}
{Ness} M.~K.,  {Johnston} K.~V.,  {Blancato} K.,  {Rix} H.~W.,  {Beane} A.,
  {Bird} J.~C.,   {Hawkins} K.,  2019, \mn@doi [\apj]
  {10.3847/1538-4357/ab3e3c}, \href
  {https://ui.adsabs.harvard.edu/abs/2019ApJ...883..177N} {883, 177}

\bibitem[\protect\citeauthoryear{{Nielsen} et~al.,}{{Nielsen}
  et~al.}{2021}]{Nielsen:2021}
{Nielsen} M.~B.,  et~al., 2021, \mn@doi [\aj] {10.3847/1538-3881/abcd39}, \href
  {https://ui.adsabs.harvard.edu/abs/2021AJ....161...62N} {161, 62}

\bibitem[\protect\citeauthoryear{{Nomoto}, {Kobayashi}  \& {Tominaga}}{{Nomoto}
  et~al.}{2013}]{Nomoto:2013aa}
{Nomoto} K.,  {Kobayashi} C.,   {Tominaga} N.,  2013, \mn@doi [\araa]
  {10.1146/annurev-astro-082812-140956}, \href
  {http://adsabs.harvard.edu/abs/2013ARA%26A..51..457N} {51, 457}

\bibitem[\protect\citeauthoryear{{Nordlander} \& {Lind}}{{Nordlander} \&
  {Lind}}{2017}]{Nordlander:2017}
{Nordlander} T.,  {Lind} K.,  2017, \mn@doi [\aap]
  {10.1051/0004-6361/201730427}, \href
  {https://ui.adsabs.harvard.edu/abs/2017A&A...607A..75N} {607, A75}

\bibitem[\protect\citeauthoryear{{Osorio} \& {Barklem}}{{Osorio} \&
  {Barklem}}{2016}]{Osorio:2016}
{Osorio} Y.,  {Barklem} P.~S.,  2016, \mn@doi [\aap]
  {10.1051/0004-6361/201526958}, \href
  {https://ui.adsabs.harvard.edu/abs/2016A&A...586A.120O} {586, A120}

\bibitem[\protect\citeauthoryear{{Osorio}, {Barklem}, {Lind}, {Belyaev},
  {Spielfiedel}, {Guitou}  \& {Feautrier}}{{Osorio} et~al.}{2015}]{Osorio:2015}
{Osorio} Y.,  {Barklem} P.~S.,  {Lind} K.,  {Belyaev} A.~K.,  {Spielfiedel} A.,
   {Guitou} M.,   {Feautrier} N.,  2015, \mn@doi [\aap]
  {10.1051/0004-6361/201525846}, \href
  {https://ui.adsabs.harvard.edu/abs/2015A&A...579A..53O} {579, A53}

\bibitem[\protect\citeauthoryear{{Pinsonneault} et~al.,}{{Pinsonneault}
  et~al.}{2014}]{Pinsonneault:2014aa}
{Pinsonneault} M.~H.,  et~al., 2014, \mn@doi [\apjs]
  {10.1088/0067-0049/215/2/19}, \href
  {http://adsabs.harvard.edu/abs/2014ApJS..215...19P} {215, 19}

\bibitem[\protect\citeauthoryear{{Pinsonneault} et~al.,}{{Pinsonneault}
  et~al.}{2018}]{Pinsonneault:2018}
{Pinsonneault} M.~H.,  et~al., 2018, \mn@doi [\apjs]
  {10.3847/1538-4365/aaebfd}, \href
  {https://ui.adsabs.harvard.edu/abs/2018ApJS..239...32P} {239, 32}

\bibitem[\protect\citeauthoryear{{Queiroz} et~al.,}{{Queiroz}
  et~al.}{2018}]{Queiroz:2018}
{Queiroz} A.~B.~A.,  et~al., 2018, \mn@doi [\mnras] {10.1093/mnras/sty330},
  \href {https://ui.adsabs.harvard.edu/abs/2018MNRAS.476.2556Q} {476, 2556}

\bibitem[\protect\citeauthoryear{{Queiroz} et~al.,}{{Queiroz}
  et~al.}{2020}]{Queiroz:2020}
{Queiroz} A.~B.~A.,  et~al., 2020, \mn@doi [\aap]
  {10.1051/0004-6361/201937364}, \href
  {https://ui.adsabs.harvard.edu/abs/2020A&A...638A..76Q} {638, A76}

\bibitem[\protect\citeauthoryear{{Reimers}}{{Reimers}}{1975}]{Reimers:1975}
{Reimers} D.,  1975, Memoires of the Societe Royale des Sciences de Liege,
  \href {https://ui.adsabs.harvard.edu/abs/1975MSRSL...8..369R} {8, 369}

\bibitem[\protect\citeauthoryear{{Rodrigues} et~al.,}{{Rodrigues}
  et~al.}{2014}]{Rodrigues:2014}
{Rodrigues} T.~S.,  et~al., 2014, \mn@doi [\mnras] {10.1093/mnras/stu1907},
  \href {https://ui.adsabs.harvard.edu/abs/2014MNRAS.445.2758R} {445, 2758}

\bibitem[\protect\citeauthoryear{{Rodrigues} et~al.,}{{Rodrigues}
  et~al.}{2017}]{Rodrigues:2017}
{Rodrigues} T.~S.,  et~al., 2017, \mn@doi [\mnras] {10.1093/mnras/stx120},
  \href {https://ui.adsabs.harvard.edu/abs/2017MNRAS.467.1433R} {467, 1433}

\bibitem[\protect\citeauthoryear{{Ro{\v{s}}kar}, {Debattista}, {Quinn},
  {Stinson}  \& {Wadsley}}{{Ro{\v{s}}kar} et~al.}{2008}]{Roskar:2008}
{Ro{\v{s}}kar} R.,  {Debattista} V.~P.,  {Quinn} T.~R.,  {Stinson} G.~S.,
  {Wadsley} J.,  2008, \mn@doi [\apjl] {10.1086/592231}, \href
  {https://ui.adsabs.harvard.edu/abs/2008ApJ...684L..79R} {684, L79}

\bibitem[\protect\citeauthoryear{{Salaris}, {Chieffi}  \&
  {Straniero}}{{Salaris} et~al.}{1993}]{Salaris:1993}
{Salaris} M.,  {Chieffi} A.,   {Straniero} O.,  1993, \mn@doi [\apj]
  {10.1086/173105}, \href
  {https://ui.adsabs.harvard.edu/abs/1993ApJ...414..580S} {414, 580}

\bibitem[\protect\citeauthoryear{{Sch{\"o}nrich} \& {Binney}}{{Sch{\"o}nrich}
  \& {Binney}}{2009}]{Schonrich:2009aa}
{Sch{\"o}nrich} R.,  {Binney} J.,  2009, \mn@doi [\mnras]
  {10.1111/j.1365-2966.2009.14750.x}, \href
  {http://adsabs.harvard.edu/abs/2009MNRAS.396..203S} {396, 203}

\bibitem[\protect\citeauthoryear{{Sch{\"o}nrich}, {Binney}  \&
  {Dehnen}}{{Sch{\"o}nrich} et~al.}{2010}]{Schonrich:2010}
{Sch{\"o}nrich} R.,  {Binney} J.,   {Dehnen} W.,  2010, \mn@doi [\mnras]
  {10.1111/j.1365-2966.2010.16253.x}, \href
  {https://ui.adsabs.harvard.edu/abs/2010MNRAS.403.1829S} {403, 1829}

\bibitem[\protect\citeauthoryear{{Sellwood} \& {Binney}}{{Sellwood} \&
  {Binney}}{2002}]{Sellwood:2002}
{Sellwood} J.~A.,  {Binney} J.~J.,  2002, \mn@doi [\mnras]
  {10.1046/j.1365-8711.2002.05806.x}, \href
  {https://ui.adsabs.harvard.edu/abs/2002MNRAS.336..785S} {336, 785}

\bibitem[\protect\citeauthoryear{{Serenelli} et~al.,}{{Serenelli}
  et~al.}{2017}]{Serenelli:2017}
{Serenelli} A.,  et~al., 2017, \mn@doi [\apjs] {10.3847/1538-4365/aa97df},
  \href {https://ui.adsabs.harvard.edu/abs/2017ApJS..233...23S} {233, 23}

\bibitem[\protect\citeauthoryear{{Silva Aguirre} et~al.,}{{Silva Aguirre}
  et~al.}{2015}]{SilvaAguirre:2015}
{Silva Aguirre} V.,  et~al., 2015, \mn@doi [\mnras] {10.1093/mnras/stv1388},
  \href {https://ui.adsabs.harvard.edu/abs/2015MNRAS.452.2127S} {452, 2127}

\bibitem[\protect\citeauthoryear{{Silva Aguirre} et~al.,}{{Silva Aguirre}
  et~al.}{2017}]{SilvaAguirre:2017}
{Silva Aguirre} V.,  et~al., 2017, \mn@doi [\apj]
  {10.3847/1538-4357/835/2/173}, \href
  {https://ui.adsabs.harvard.edu/abs/2017ApJ...835..173S} {835, 173}

\bibitem[\protect\citeauthoryear{{Silva Aguirre} et~al.,}{{Silva Aguirre}
  et~al.}{2018}]{SilvaAguirre:2018}
{Silva Aguirre} V.,  et~al., 2018, \mn@doi [\mnras] {10.1093/mnras/sty150},
  \href {https://ui.adsabs.harvard.edu/abs/2018MNRAS.475.5487S} {475, 5487}

\bibitem[\protect\citeauthoryear{{Sneden}}{{Sneden}}{1973}]{Sneden:1973phd}
{Sneden} C.~A.,  1973, PhD thesis, THE UNIVERSITY OF TEXAS AT AUSTIN.

\bibitem[\protect\citeauthoryear{{Spina}, {Mel{\'e}ndez}, {Karakas},
  {Ram{\'{\i}}rez}, {Monroe}, {Asplund}  \& {Yong}}{{Spina}
  et~al.}{2016}]{Spina:2016aa}
{Spina} L.,  {Mel{\'e}ndez} J.,  {Karakas} A.~I.,  {Ram{\'{\i}}rez} I.,
  {Monroe} T.~R.,  {Asplund} M.,   {Yong} D.,  2016, \mn@doi [\aap]
  {10.1051/0004-6361/201628557}, \href
  {http://adsabs.harvard.edu/abs/2016A%26A...593A.125S} {593, A125}

\bibitem[\protect\citeauthoryear{{Spitoni}, {Silva Aguirre}, {Matteucci},
  {Calura}  \& {Grisoni}}{{Spitoni} et~al.}{2019}]{Spitoni:2019}
{Spitoni} E.,  {Silva Aguirre} V.,  {Matteucci} F.,  {Calura} F.,   {Grisoni}
  V.,  2019, \mn@doi [\aap] {10.1051/0004-6361/201834188}, \href
  {https://ui.adsabs.harvard.edu/abs/2019A&A...623A..60S} {623, A60}

\bibitem[\protect\citeauthoryear{{Stello} et~al.,}{{Stello}
  et~al.}{2011}]{Stello:2011}
{Stello} D.,  et~al., 2011, \mn@doi [\apj] {10.1088/0004-637X/739/1/13}, \href
  {https://ui.adsabs.harvard.edu/abs/2011ApJ...739...13S} {739, 13}

\bibitem[\protect\citeauthoryear{{Stello} et~al.,}{{Stello}
  et~al.}{2015}]{Stello:2015}
{Stello} D.,  et~al., 2015, \mn@doi [\apjl] {10.1088/2041-8205/809/1/L3}, \href
  {https://ui.adsabs.harvard.edu/abs/2015ApJ...809L...3S} {809, L3}

\bibitem[\protect\citeauthoryear{{Tailo} et~al.,}{{Tailo}
  et~al.}{2020}]{Tailo:2020}
{Tailo} M.,  et~al., 2020, \mn@doi [\mnras] {10.1093/mnras/staa2639}, \href
  {https://ui.adsabs.harvard.edu/abs/2020MNRAS.498.5745T} {498, 5745}

\bibitem[\protect\citeauthoryear{{Tatischeff} \& {Gabici}}{{Tatischeff} \&
  {Gabici}}{2018}]{Tatischeff:2018}
{Tatischeff} V.,  {Gabici} S.,  2018, \mn@doi [Annual Review of Nuclear and
  Particle Science] {10.1146/annurev-nucl-101917-021151}, \href
  {https://ui.adsabs.harvard.edu/abs/2018ARNPS..68..377T} {68, 377}

\bibitem[\protect\citeauthoryear{{Ting}, {Hawkins}  \& {Rix}}{{Ting}
  et~al.}{2018}]{Ting:2018}
{Ting} Y.-S.,  {Hawkins} K.,   {Rix} H.-W.,  2018, \mn@doi [\apjl]
  {10.3847/2041-8213/aabf8e}, \href
  {https://ui.adsabs.harvard.edu/abs/2018ApJ...858L...7T} {858, L7}

\bibitem[\protect\citeauthoryear{{Trevisan}, {Barbuy}, {Eriksson},
  {Gustafsson}, {Grenon}  \& {Pomp{\'e}ia}}{{Trevisan}
  et~al.}{2011}]{Trevisan:2011}
{Trevisan} M.,  {Barbuy} B.,  {Eriksson} K.,  {Gustafsson} B.,  {Grenon} M.,
  {Pomp{\'e}ia} L.,  2011, \mn@doi [\aap] {10.1051/0004-6361/201016056}, \href
  {https://ui.adsabs.harvard.edu/abs/2011A&A...535A..42T} {535, A42}

\bibitem[\protect\citeauthoryear{Virtanen et~al.,}{Virtanen
  et~al.}{2020}]{scipy:2020}
Virtanen P.,  et~al., 2020, \mn@doi [Nature Methods]
  {10.1038/s41592-019-0686-2}, \href {https://rdcu.be/b08Wh} {17, 261}

\bibitem[\protect\citeauthoryear{{Vogt} et~al.,}{{Vogt}
  et~al.}{1994}]{Vogt:1994}
{Vogt} S.~S.,  et~al., 1994, in {Crawford} D.~L.,  {Craine} E.~R.,  eds,
  Society of Photo-Optical Instrumentation Engineers (SPIE) Conference Series
  Vol. 2198, Instrumentation in Astronomy VIII. p.~362,
  \mn@doi{10.1117/12.176725}

\bibitem[\protect\citeauthoryear{{Voronov}, {Yakovleva}  \&
  {Belyaev}}{{Voronov} et~al.}{2022}]{Voronov:2022}
{Voronov} Y.~V.,  {Yakovleva} S.~A.,   {Belyaev} A.~K.,  2022, \mn@doi [\apj]
  {10.3847/1538-4357/ac46fd}, \href
  {https://ui.adsabs.harvard.edu/abs/2022ApJ...926..173V} {926, 173}

\bibitem[\protect\citeauthoryear{{Wielen}, {Fuchs}  \& {Dettbarn}}{{Wielen}
  et~al.}{1996}]{Wielen:1996}
{Wielen} R.,  {Fuchs} B.,   {Dettbarn} C.,  1996, \aap, \href
  {https://ui.adsabs.harvard.edu/abs/1996A&A...314..438W} {314, 438}

\bibitem[\protect\citeauthoryear{{Xiang} et~al.,}{{Xiang}
  et~al.}{2019}]{Xiang:2019}
{Xiang} M.,  et~al., 2019, \mn@doi [\apjs] {10.3847/1538-4365/ab5364}, \href
  {https://ui.adsabs.harvard.edu/abs/2019ApJS..245...34X} {245, 34}

\bibitem[\protect\citeauthoryear{{Yong} et~al.,}{{Yong}
  et~al.}{2014}]{Yong:2014}
{Yong} D.,  et~al., 2014, \mn@doi [\mnras] {10.1093/mnras/stu118}, \href
  {https://ui.adsabs.harvard.edu/abs/2014MNRAS.439.2638Y} {439, 2638}

\bibitem[\protect\citeauthoryear{{Yong} et~al.,}{{Yong}
  et~al.}{2016}]{Yong:2016aa}
{Yong} D.,  et~al., 2016, \mn@doi [\mnras] {10.1093/mnras/stw676}, \href
  {http://adsabs.harvard.edu/abs/2016MNRAS.459..487Y} {459, 487}

\bibitem[\protect\citeauthoryear{{Yu}, {Huber}, {Bedding}, {Stello}, {Hon},
  {Murphy}  \& {Khanna}}{{Yu} et~al.}{2018}]{Yu:2018}
{Yu} J.,  {Huber} D.,  {Bedding} T.~R.,  {Stello} D.,  {Hon} M.,  {Murphy}
  S.~J.,   {Khanna} S.,  2018, \mn@doi [\apjs] {10.3847/1538-4365/aaaf74},
  \href {https://ui.adsabs.harvard.edu/abs/2018ApJS..236...42Y} {236, 42}

\bibitem[\protect\citeauthoryear{{Zhang} et~al.,}{{Zhang}
  et~al.}{2021}]{Zhang:2021}
{Zhang} M.,  et~al., 2021, \mn@doi [\apj] {10.3847/1538-4357/ac22a5}, \href
  {https://ui.adsabs.harvard.edu/abs/2021ApJ...922..145Z} {922, 145}

\bibitem[\protect\citeauthoryear{{Zinn}, {Pinsonneault}, {Huber}  \&
  {Stello}}{{Zinn} et~al.}{2019}]{Zinn2019}
{Zinn} J.~C.,  {Pinsonneault} M.~H.,  {Huber} D.,   {Stello} D.,  2019, \mn@doi
  [\apj] {10.3847/1538-4357/ab1f66}, \href
  {https://ui.adsabs.harvard.edu/abs/2019ApJ...878..136Z} {878, 136}

\bibitem[\protect\citeauthoryear{{Zinn} et~al.,}{{Zinn}
  et~al.}{2022}]{Zinn:2022}
{Zinn} J.~C.,  et~al., 2022, \mn@doi [\apj] {10.3847/1538-4357/ac2c83}, \href
  {https://ui.adsabs.harvard.edu/abs/2022ApJ...926..191Z} {926, 191}

\bibitem[\protect\citeauthoryear{{da Silva} et~al.,}{{da Silva}
  et~al.}{2006}]{daSilva:2006}
{da Silva} L.,  et~al., 2006, \mn@doi [\aap] {10.1051/0004-6361:20065105},
  \href {https://ui.adsabs.harvard.edu/abs/2006A&A...458..609D} {458, 609}

\makeatother
\end{thebibliography}




\appendix

\section{IRFM calculations}

\begin{table}
	\centering
	\caption{Bolometric fluxes and angular diameters calculated with the IRFM. Typical uncertainties are of the order of 2-4\%.}
	\label{tab:ap_irfm}
	\begin{tabular}{lrr}
		\hline
		    Star &  Bolometric Flux & Angular Diameter \\
	             &  erg~s$^{-1}$~cm$^{-2}$ &     (mas) \\
        \hline
             KIC\,2845610 & 5.894e-09 & 0.153 \\
             KIC\,3455760 & 1.687e-09 & 0.101 \\
             KIC\,3833399 & 6.581e-09 & 0.194 \\
             KIC\,5512910 & 2.219e-10 & 0.034 \\
             KIC\,5707338 & 2.640e-09 & 0.110 \\
             KIC\,6605673 & 9.698e-10 & 0.048 \\
             KIC\,6634419 & 1.250e-09 & 0.068 \\
             KIC\,6936796 & 1.165e-09 & 0.092 \\
             KIC\,6940126 & 1.861e-10 & 0.032 \\
             KIC\,7595155 & 6.452e-10 & 0.067 \\
             KIC\,8145677 & 7.152e-10 & 0.056 \\
             KIC\,9002884 & 9.384e-10 & 0.094 \\
             KIC\,9266192 & 6.827e-09 & 0.173 \\
             KIC\,9761625 & 8.788e-10 & 0.083 \\
            KIC\,10525475 & 1.547e-09 & 0.092 \\
            KIC\,11823838 & 1.273e-09 & 0.081 \\
        \hline
	\end{tabular}
\end{table}


\bsp	
\label{lastpage}
\end{document}